\documentclass[fleqn,usenatbib]{mnras}

\usepackage{newtxtext,newtxmath}

\usepackage{amsmath}	
\usepackage{amssymb}	
\usepackage{multicol}        
\usepackage{bm}		
\usepackage{pdflscape}	
\usepackage[T1]{fontenc}
\usepackage{ae,aecompl}
\usepackage{comment}
\usepackage{caption}
\usepackage{subcaption}
\usepackage[utf8]{inputenc}
\usepackage{float}
\usepackage{graphicx}	
\usepackage{amsmath}	
\usepackage{amssymb}	

\title[Comparing Foreground Removal Techniques]{Comparing Foreground Removal Techniques for Recovery of the LOFAR-EoR 21cm Power Spectrum}
\author[Hothi et al.]{Ian Hothi$^{1}$\thanks{E-mail: i.hothi18@imperial.ac.uk},
Emma Chapman$^{1}$,
Jonathan R. Pritchard$^{1}$,
F. G. Mertens$^{2,3}$,
\newauthor 
L.V.E Koopmans$^{2}$,
B. Ciardi$^{4}$,
B.K. Gehlot$^{5}$,
R. Ghara$^{6,7}$,
A. Ghosh$^{8}$,
S. K. Giri$^{9}$,
\newauthor
I. T. Iliev$^{10}$,
V. Jeli\'{c}$^{11}$,
S. Zaroubi$^{2,6,7}$
\\
$^{1}$Department of Physics, Blackett Laboratory, Imperial College London, SW7 2AZ, U.K
\\
$^{2}$Kapteyn Astronomical Institute, University of Groningen, P. O. Box 800 9700 AV Groningen, The Netherlands
\\
$^{3}$LERMA, Observatoire de Paris, PSL Research University, CNRS, Sorbonne Universit{\'e}, F-75014 Paris, France
\\
$^{4}$Max Planck Institute for Astrophysics, Karl-Schwarzschild-Str. 1, 85741 Garching, Germany
\\
$^{5}$School of Earth and Space Exploration, Arizona State University, 781 Terrace Mall, Tempe, AZ 85287, U.S.A.
\\
$^{6}$Department of Natural Sciences, The Open University of Israel, 1 University Road, PO Box 808, Ra'anana 4353701, Israel 
\\
$^{7}$Department of Physics, Technion, Haifa 32000, Israel
\\
$^{8}$Department of Physics, Banwarilal Bhalotia College, GT Rd, Ushagram, Asansol, West Bengal 713303, India
\\
$^{9}$Institute for Computational Science, University of Zurich, Winterthurerstrasse 190, 8057 Zurich, Switzerland
\\
$^{10}$Astronomy Centre, Department of Physics and Astronomy, Pevensey II Building, University of Sussex, Brighton BN1 9QH, U.K.
\\
$^{11}$Ru{\dj}er Bo\v{s}kovi\'c Institute, Bijeni\v{c}ka cesta 54, 10 000 Zagreb, Croatia
}
\date{Accepted 2020 November 02. Received 2020 November 02; in original form 2020 October 09}

\pubyear{2020}

\begin{document}
\label{firstpage}
\pagerange{\pageref{firstpage}--\pageref{lastpage}}
\maketitle

\begin{abstract}
We compare various foreground removal techniques that are being utilised to remove bright foregrounds in various experiments aiming to detect the redshifted 21cm signal of neutral hydrogen from the Epoch of Reionization. In this work, we test the performance of removal techniques (FastICA, GMCA, and GPR) on 10 nights of LOFAR data and investigate the possibility of recovering the latest upper limit on the 21cm signal. Interestingly, we find that GMCA and FastICA reproduce the most recent 2$\sigma$ upper limit of $\Delta^2_{21} <$ (73)$^2$ mK$^2$ at $k=0.075~ h \mathrm{cMpc}^{-1}$, which resulted from the application of GPR. We also find that FastICA and GMCA begin to deviate from the noise-limit at \textit{k}-scales larger than $\sim 0.1 ~h \mathrm{cMpc}^{-1}$. We then replicate the data via simulations to see the source of FastICA and GMCA's limitations, by testing them against various instrumental effects. We find that no single instrumental effect, such as primary beam effects or mode-mixing, can explain the poorer recovery by FastICA and GMCA at larger \textit{k}-scales. We then test scale-independence of FastICA and GMCA, and find that lower \textit{k}-scales can be modelled by a smaller number of independent components. For larger scales ($k \gtrsim 0.1~h \mathrm{cMpc}^{-1}$), more independent components are needed to fit the foregrounds. We conclude that, the current usage of GPR by the LOFAR collaboration is the appropriate removal technique. It is both robust and less prone to overfitting, with future improvements to GPR's fitting optimisation to yield deeper limits.
\end{abstract}




\section{Introduction}
The early Universe was predominantly neutral with sources of radiation yet to form and was aptly named the `Dark Ages' \citep{Jon_Review}. The first sources of ionizing radiation formed 80 million years after the Big Bang, from the inhomogeneities seeded by inflation - marking the Cosmic Dawn \citep{EDGES}. These sources ionize their neutral surroundings producing ionized bubbles. As these bubbles grow, and more sources form, they begin to merge, slowly ionizing the neutral Intergalactic Medium (IGM) - the onset of the Epoch of Reionization (EoR) \citep{2005:CiardiFerrara,2010:MoralesWyithe,2016Furlanetto}. The EoR was the last major phase change in the Universe's history. 

Indirect observations of the EoR have provided constraints. Quasars strongly emit Lyman-$\alpha$ photons and, as they travel towards us, the photons are redshifted. Any photons that are blue-ward of the Ly-$\alpha$ photon, at a particular redshift, will be redshifted such that it corresponds to the Ly-$\alpha$ photons. If neutral hydrogen is present at this redshift, the Ly-$\alpha$ photon will be absorbed. This leads to the spectra of quasars having a large region of purely absorption - the Gunn-Peterson trough \citep{Zaroubi_Review}. Looking at the prominence of this feature, at progressively lower redshifts the trough starts to become incomplete and there is no longer a sharp rise that marks the end of the trough. This might suggest that the IGM is largely ionized and nearing full ionization by z $\sim$ 6, leaving only a few neutral islands \citep{GPTroughFan}. In the spectra of high-redshift quasars, their Lyman-$\alpha$ damping absorption have been used to constrain the EoR, as having not been completed by a redshift of $z \sim$ 7 \citep{2017MNRAS.466.4239G,2019MNRAS.484.5094G}. Reionization leads to an increase of free electrons in the IGM, and CMB photons will scatter off of these free electrons. The scattering of these photons by the free electrons allows us to calculate an associated optical depth to the CMB, $\tau$ (the total integrated optical depth to reionization), which indicates the amount of scattering that has occurred along the line of sight. This scattering leads to a scale-independent power suppression at scales $l\geq10$ by a factor of e$^{2\tau}$. 
\citet{planck2018} found $\tau$ = 0.054 $\pm$ 0.007. Using this, one can model the evolution of the neutral fraction and infer the bulk of reionization must have occurred at $z \lesssim  14 $ \citep{PlanckEoR16}. Combining these indirect observations, we can loosely constrain reionization to occur between redshifts 14 to 6 \citep{Ade}.

The foremost probe of the EoR is the 21cm hyperfine forbidden line transition from neutral hydrogen \citep{Wouthuysen1952,Field58,shaver}. The 21cm signal is a powerful probe, as reionization can imprint features in the 21cm power spectrum telling us the topology of reionization which itself is sensitive to the properties of ionizing sources \citep{2006Furlanetto,2016Furlanetto}. 

Though it has thus far eluded detection, the EoR is likely to be directly observed for the first time by radio telescopes currently taking data, e.g. Low Frequency Array (LOFAR)\footnote{http://www.lofar.org/} \citep{LOFAROVERVIEW}; Giant Metrewave Radio Telescope (GMRT)\footnote{http://gmrt.ncra.tifr.res.in/} \citep{GMRT1,GMRT2}; Murchison Widefield Array (MWA)\footnote{http://www.mwatelescope.org/} \citep{MWAPaper}; Precision Array to Probe the Epoch of Reionization (PAPER)\footnote{http://eor.berkeley.edu} \citep{PAPER}; the Hydrogen Epoch of Reionization Array (HERA)\footnote{http://reionization.org} \citep{HERA}, and at lower frequencies, the new extension in Nan\c{c}ay upgrading LOFAR (NeNuFar)\footnote{https://nenufar.obs-nancay.fr/en/homepage-en/} \citep{NenuFAR}; the Owens Valley Long Wavelength Array (OVRO-LWA)\footnote{http://www.tauceti.caltech.edu/LWA/}, and  Amsterdam-ASTRON Radio Transients Facility and Analysis Center (AARTFAAC)\footnote{http://aartfaac.org} \citep{AARFAAC_16,gehlot2019}. These are interferometric experiments that aim to measure the statistical fluctuations of the 21cm signal. 

There also exist single antenna experiments that aim to measure the sky-averaged global 21cm signal. These include, the Global HydrOgen ReioNization Signal (BIGHORNS)\footnote{http://www.mwatelescope.org/telescope/external/bighorns} \citep{BIGHORNS}; the Sonda Cosmol\'ogica de las Islas para la Detecci\'{o}n deHidr\'{o}geno Neutro (SCI-HI) \citep{SCI-HI}; the Experiment to Detect the Global Epoch of Reionization Signature (EDGES)\footnote{https://loco.lab.asu.edu/edges/} \citep{EDGES}; Shaped Antenna measurement of the background RAdio Spectrum 2 (SARAS 2)\footnote{http://www.rri.res.in/DISTORTION/saras.html}  \citep{SARAS2}; Large-aperture Experiment to Detect the Dark Age (LEDA) \footnote{http://www.tauceti.caltech.edu/leda/}  \citep{LEDA}, and Probing Radio Intensity at High-Z (PRIZM)  \citep{PRIZM}. \citet{EDGES} has a tentative detection of this global signal, with the absorption depth and width, currently, unexplained. Though there have been claims that the results could be explained by exotic physics \citep{DMSCATTER}, potential unaccounted for systematics \citep{EDGESYS}, foreground polarizations \citep{EDGESPOL}, or modelling \citep{PoorModellingEDGES}. 

The biggest obstacle preventing the detection of the EoR by radio interferometers are the foregrounds that swamp the observed signal. These foregrounds are several orders of magnitude stronger than the 21cm signal we wish to observe \citep{shaver,OhMac,Matteo04,Jov08}. These foregrounds are modelled to be smooth (in the frequency domain) and this was exploited by the early foreground removal techniques, e.g. polynomial fitting \citep{Jov08}. This smooth nature of the foregrounds spectra is likely to deviate when met with instrumental effects, and so non-parametric techniques, such as blind source separation (BSS), are often favoured as they do not explicitly state the form of foregrounds a-priori  \citep{FASTICA12a,GMCA13}. More recently Bayesian techniques, such as Gaussian process regression (GPR) \citep{GPR}, have been looked at as an optimised fitting tool with the priors set by the various components of the observed data, such as the foregrounds and cosmological signal. Several machine learning methods have shown promising signs for the role of machine learning in further study of the EoR \citep{2019:HassanLiuKohn,2019:LiXuMa,2019:LaPlanteNtampaka,2020:MangenaHassanSantos}.

The aforementioned experiments have already started to provide upper limits on the 21cm power spectrum. GMRT \citep{GMRT2} provided an early 2$\sigma$ upper limit, with 40 hours worth of data, at a redshift of 8.6 with $\Delta^2 \simeq$  (248 mK)$^2$ at \textit{k} = 0.50 \textit{h}cMpc$^{-1}$. Here they used Singular Value Decomposition (SVD) on each baseline, individually, with visibilities represented in a matrix by time and frequency. SVD isolates the smooth foreground modes for each baseline, removing these modes - the residual from which the power spectrum is calculated on. 

MWA \citep{BarryMWAUpper}, using 21 hours worth of data in \citet{MWA_upper}'s MWA observing run to produce a limit of $\Delta^2 \simeq$ (62.4 mK)$^2$ at \textit{k} = 0.20 \textit{h}cMpc$^{-1}$, at a redshift of 7. \citet{LiUpperMWA}, produced a 2$\sigma$ upper limit of $\Delta^2 \simeq$ (49 mK)$^2$ at \textit{k} = 0.59 \textit{h}cMpc$^{-1}$, at a redshift of 6.5. More recently, \citet{CATHUPPER} used 110 hours of data to produce an upper limit of $\Delta^2 \simeq$ (43 mK)$^2$ at \textit{k} = 0.14 \textit{h}cMpc$^{-1}$ at a redshift of 6.5. Here they used a model for the point sources remove within their primary beam, before weighting their power spectrum estimation, to reduce the impact of foregrounds.

PAPER \citep{PAPERUPPERNEW}, Fourier transformed each baseline spectrum into the delay domain and then filtered out foreground modes in this domain \citep{PAPERANAL}. This produced the upper limit, between 0.3 $<$ \textit{k} $<$ 0.6 \textit{h}cMpc$^{-1}$, of $\Delta^2 \simeq$ (200 mK)$^2$ at redshift 8.37.

LOFAR's first set of upper limits, from their high-band antenna (HBA), came in \citet{UpperLim_HBA}. They produced a 2$\sigma$ upper limit of $\Delta^2 \simeq$ (79.6 mK)$^2$ at \textit{k} = 0.053 \textit{h}cMpc$^{-1}$ in the redshift range 9.6 to 10.6. Here they produced an upper limit using Generalised Morphological Component Analysis (GMCA) \citep{GMCA13}, as its primary removal technique. More recently, LOFAR have used GPR for upper limits,  \citep{GPR,LOFAR_UPPERLIMIT_GPR}, the application of which resulted in the recent 2$\sigma$ upper limit of  $\Delta^2 <$ (73 mK)$^2$ at \textit{k} = 0.075 \textit{h}cMpc$^{-1}$ at a redshift of 9.1, using 141 hours worth of LOFAR data \citep{MertensUpper2020}. This recent upper limit has already been used to provide constraints on the intergalactic medium, excess radio background, and the neutral fraction at  a redshift of 9.1 \citep{2020:GharaGiriMellema,2020:MondalFialkovFling,2020:GreigMesingerKoopmans}.

The goal of this paper is to assess the foreground removal techniques' pros and cons, by applying them to the 10 nights of LOFAR data as well as simulated data - with both having an injected simulated 21cm signal. The paper is written as follows. In Section \ref{Removal}, we discuss the foreground removal techniques used in this paper, before moving on to Section \ref{Data}, which describes the data. We then perform a full analysis of the power spectra results of the removal techniques in Section \ref{result}, followed by an investigation into the results. We summarise our results and conclude the paper with a discussion in Section \ref{discussion}.

\section{Foreground Removal Techniques}

In this section, we will provide an overview of the foreground removal techniques used in this paper, FastICA, GMCA, and GPR. We will detail how they model the foregrounds and recover the 21cm signal.

We had also attempted to apply another non-parametric removal technique, Wp smoothing, to our data \citep{WpSmooth}. Whilst it worked on the simulated data without any noise or instrumentation effects, it was unable to fit a function to the observed data. We believe that the chromaticity within the data, along with the small amount of frequencies study, result in the smoothing algorithms being unable to fit a function.
\label{Removal}
\subsection{FastICA}
FastICA is a non-parametric technique that aims to model the foregrounds by assuming they are statistically independent in the mixed signal we observe \citep{FastICA_Book}. FastICA has been used in the medical field for functional magnetic resonance imaging (fMRI) analysis \citep{fMRI03,fMRI16}, as well as decomposing spinal cord signals \citep{SpinalCord}. Early uses of FastICA for foreground removal, include its application to CMB data. Both for recovering the  spatial pattern and the frequency scalings of foregrounds, as well as removing foreground to detect the CMB \citep{2002FastICACMB,2003FastICACMB}. FastICA has also been used for foreground removal in HI mapping \citep{FastICA_App}. The use of this non-parametric foreground reduction method on 21cm data, was outlined by \citet{FASTICA12a}. The data is modelled as:
\begin{equation}\label{eq:FASTICAMAIN}
\textbf{x} = \textbf{As + n},
\end{equation}
where \textbf{x} is the vector representation of the observed signal, \textbf{s} is a vector containing independent components, and \textbf{A} is the mixing matrix we wish to find. The algorithm does not attempt to fit the additive Gaussian noise term, \textbf{n}. The number of expected independent components comprising the foregrounds, is given to the algorithm, and FastICA attempts to reconstruct this number. To solve for \textbf{s} we consider the following linear transform:
\begin{equation}\label{eq:FASTICAS}
\textbf{s} = \textbf{Wx},
\end{equation}
with \textbf{W} being a constant weight matrix for FastICA to determine. It finds \textbf{W} by maximising the non-Gaussianity of equation \eqref{eq:FASTICAS}, quantifying non-Gaussianity by negentropy - this uses the principle that the greater the number of independent components in the set, the more Gaussian the distribution of the set will be (Central Limit Theorem).

Let us consider the example for finding a single component of \textbf{s}. We define,

\begin{equation}\label{eq:FASTICA_y}
y = \textbf{w$^T$x}.
\end{equation}

Here, \textbf{w} is a row from the inverse-matrix of \textbf{A}. As such, \textit{y} represents a single component (row) of \textbf{s}. We first define the entropy of a variable as,

\begin{equation}\label{eq:entropy}
H(y) = - \sum_i P(y=a_i)\: \rm{log}\,P(y=a_i),
\end{equation}
where $a_i$ is a possible value of y. Following equation \eqref{eq:entropy}, we define negentropy as,

\begin{equation}\label{eq:negentropy}
J(y) = H(y_{gauss}) - H(y).
\end{equation}

With $y_{gauss}$ being a random Gaussian variable with an identical covariance matrix as \textit{y}. From the central limit theorem, by maximising the negentropy of \textit{y}, and hence the non-Gaussianity, we retrieve a single independent component of \textbf{s}.

By adjusting the weighting matrix, using negentropy as a measure of equation \eqref{eq:FASTICAS}'s non-Gaussianity, one can find the independent variables corresponding to the foregrounds. 

Once \textbf{W} is found, we can find \textbf{A} as the inverse of \textbf{W} (\textbf{A} = \textbf{W}$^{-1}$), and \textbf{s} can be found via equation \eqref{eq:FASTICAS}. Once FastICA has calculated \textbf{s}, one can use the mixing matrix to reconstruct the foreground signal, \textbf{As}, and deduct from \textbf{x}. The residual of this will be a mixture of the 21cm signal, noise and any fitting errors. 
\subsection{GMCA}
Generalised Morphological Component Analysis (GMCA) separates out the foregrounds from the signal (noise + EoR signal) by finding the sparsest components in an orthogonal wavelet basis that the foregrounds can be represented by \citep{GMCA13}. It has previously been used by CMB experiments to remove Galactic foregrounds and point sources from their maps \citep{GMCA_App}.

More recently, \citet{2020:CarucciIrfanBobin} have applied GMCA to 21cm maps and have also modelled, previously unaccounted for 'foreground', polarisation leakage. GMCA also models the data with equation \eqref{eq:FASTICAMAIN}. To solve this problem, we need to estimate \textbf{A} and \textbf{s}, where the latter represents the foreground signals which we can decompose (in the wavelet basis) as:  \textbf{s} = $\sum^T_{j=0}\textbf{s}_j$. Where, $\textbf{s}_j = \sum^T_{k=1}\alpha_j[k]$  \textbf{$\phi$}$_k$, such that \textbf{$\phi$}$_k$ is a wavelet waveform. As we are looking for the sparsest $s_j$, we define it to be sparse when a few $\alpha_j[k]$ are significantly non-zero.  The use of waveforms over the traditional Fourier basis functions is due to the latter being a localised basis, whereas waveforms are an infinite set in localised space. The objective of GMCA is to find a mixing matrix that yields, in the wavelet domain, the sparsest sources of \textbf{s}. This Lagrangian optimisation problem is expressed as:
\begin{equation}
\label{eq:GMCALAG}
\rm{min}(  \frac{1}{2} ||\textbf{x} - \textbf{A}\alpha\Phi||_F^2 + \lambda\sum^n_{j=0}||\alpha_j||_p  ).
\end{equation}
Once we have extracted both \textbf{A} and \textbf{s}, we can perform matrix multiplication on the two to find out foreground signal. Deducting \textbf{As} from \textbf{x} leaves us with the residual - which ideally will be the noise, EoR signal, and potentially fitting errors. 
\newline

When comparing parametric to non-parametric techniques, one will often find the former outperforming the latter, on simulated data without any realistic instrumentation effects. However, this is not expected to be true for real data, as non-parametric techniques are able to model/account for the imperfections in real data as nothing is assumed a-priori about the data. 

\subsection{GPR}
\label{GPR}
Gaussian process regression (GPR) on LOFAR was first used for foreground removal by \citet{GPR}\footnote{https://gitlab.com/flomertens/ps\_eor}, and has been used to produce the upper limits within \citet{LOFAR_UPPERLIMIT_GPR} and \citet{MertensUpper2020}. Within the GPR framework, one models the different components of observations, such as the astrophysical effects and instrumentation effects, and the cosmological signal, as a Gaussian process. Instrument effects can include those from the chromatic nature of the instrument, resulting in mode mixing \citep{ModeMixRef}. A Gaussian process is the joint distribution of a collection of normally distributed random variables, and it is described by its mean and covariance function. This covariance function is that between pairs of points at different frequencies, and hence smoothness, in frequency, of the function. The class of covariance functions used in this paper is the Matern class: 

\begin{equation}\label{eq:Matern Class}
\kappa_{Matern}(r) = \frac{2^{1-\nu}}{\Gamma(\nu)} \bigg(\frac{\sqrt{2\nu}r}{l} \bigg)^{\nu} K_{\nu}\bigg(\frac{\sqrt{2\nu}r}{l} \bigg)
\end{equation}

$K_{\nu}$ is the modified Bessel function of the second kind, and $l$ is the hyper-parameter of the kernel - the characteristic coherence-scale. Using different values of $\nu$, we retrieve special cases of this matrix. For example, setting $\nu$ to $\frac{1}{2}$ we get the exponential kernel.

We describe the Gaussian processes as parameterised priors in GPR, and this prior is selected such that it maximises the Bayesian evidence with the hyper-parameters of the covariance function being found via Markov chain Monte Carlo (MCMC) algorithms. 

GPR models the observed data \textbf{d} at frequency $\nu$ by a foreground, a 21-cm and a noise signal \textbf{n}:
\begin{equation}\label{eq:GPR_Model_Data}
\textbf{d} = \textbf{f}_{\rm{fg}} + \textbf{f}_{\rm{21}} + \textbf{n}.
\end{equation}
As the 21cm signal is expected to be uncorrelated on scales of the order of a few MHz, we can exploit this to separate from the foreground signal, which is expected to be  correlated (smooth) on scales of 1 MHz. We can thus separate out the covariance function of the Gaussian process into a covariance for the foregrounds, \textit{\textbf{K$_{\rm{fg}}$}}, and a covariance for the 21cm signal, \textit{\textbf{K$_{\rm{21}}$}}:
\begin{equation}\label{eq:Cov_Decomp}
\textbf{K} = \textit{\textbf{K}}_{\rm{fg}} + \textit{\textbf{K}}_{21}.
\end{equation}
The foreground covariance can be decomposed into two terms: The first is the intrinsic foreground \textit{\textbf{K}}$_{\rm{int}}$, which is the expected smooth/coherent foreground (over large frequency scales). The second component is for the instrumental effects, such as mode-mixing, and is coherent on a smaller frequency range. As the 21-cm signal is faint compared to the foregrounds and the noise, the inclusion of the 21cm signals covariance, \textit{\textbf{K}}$_{21}$, with appropriate priors (uniform or a more informative gamma distribution prior) in the total covariance model \textit{\textbf{K}} ensures that we don't fit away the 21-cm signal part when we optimise the maximum likelihood estimation. \citet{GPR} found the estimated coherence scale was significantly less biased, especially for a reference simulation with low signal to noise for 21-cm signal.

The joint probability distribution for the observed data \textbf{d} and function values \textbf{f}$_{\rm{fg}}$ of the foreground model at a given frequency $\nu$ is then given by,
\begin{equation}\label{eq:JPD}
\Big[\frac{\textbf{d}}{\textbf{f}_{\rm{fg}}}\Big] \sim \mathcal{N}\Big(\begin{bmatrix} 
0\\
0
\end{bmatrix},\begin{bmatrix} 
\textit{\textbf{K}}_{\rm{fg}} + \textit{\textbf{K}}_{21} + \sigma_n^2\textit{I} & \textit{\textbf{K}}_{\rm{fg}}\\
\textit{\textbf{K}}_{\rm{fg}} & \textit{\textbf{K}}_{\rm{fg}}
\end{bmatrix} \Big),
\end{equation}
where $ \sigma_{\rm{n}}^2$ is the noise variance and \textit{I} is the identity matrix. Once GPR has been performed, the foreground model is retrieved via:
\begin{equation}\label{eq:E}
E(\textbf{f}_{\rm{fg}}) = \textit{\textbf{K}}_{\rm{fg}} \Big[ \textit{\textbf{K}}_{\rm{fg}} + \textit{\textbf{K}}_{\rm{21}} + \textit{\textbf{K}}_{\rm{n}}  \Big]^{-1} \textbf{d},
\end{equation}
\begin{equation}\label{eq:Cov}
\rm{cov}(\textbf{f}_{\rm{fg}}) = \textit{\textbf{K}}_{\rm{fg}} + \textit{\textbf{K}}_{\rm{fg}} \Big[\textit{\textbf{K}}_{\rm{fg}} + \textit{\textbf{K}}_{\rm{21}} + \textit{\textbf{K}}_{\rm{n}}\Big]^{-1} \textit{\textbf{K}}_{\rm{fg}}.
\end{equation}
Where the noise covariance, \textit{\textbf{K}}$_{\rm{n}}$, is the diagonal of $\sigma_{\rm{n}}^2$: $\sigma_{\rm{n}}^2$\textit{I}. Here E(\textbf{f}$_{\rm{fg}}$) is the expectation values and \rm{cov}(\textbf{f}$_{\rm{fg}}$) is the covariance of the foregrounds. The inclusion of the \textit{\textbf{K}}$_{\rm{21}}$ kernel in the definitions allows one to marginalise over all foreground models obtained by the Gaussian process - accounting for any degeneracy between the foregrounds and 21 cm signal, and providing a proper error budget. 
Deducting E(\textbf{f}$_{\rm{fg}}$) from the original data gives the residual.

\section{Data}

In this section we will describe the data used in this paper. The first is the 10 nights of data from LOFAR that were used for the upper limit result in \citet{MertensUpper2020}. The second data set is simulated data that will be used to test the performance of the foreground removal techniques, and asses their pros and cons.

\label{Data}
\subsection{LOFAR 10 Nights Data}
The LOw-Frequency ARray (LOFAR) is an interferometric array of radio antennas with stations spread across several countries in Europe (38 stations within the Netherlands and 14 international stations). LOFAR is capable of observing the frequency range 10-240 MHz.  It is also capable of reaching an angular resolution of arcseconds across most of the 30-240 MHz range.  Stations are made up of a mixture of low band antennas - which operate from 10 MHz (just above the ionosphere cut off) to 90 MHz (just below commercial FM) - as well as high band antennas which operate between 110 MHz to 240 MHz \citep{LOFAROVERVIEW}. Most of the LOFAR stations are located in the Netherlands, with the core stations being located near the town of Exloo; chosen because of its low population and (relatively) low radio frequency interference (RFI). The core has 24 dual-stations located in a 2 km radius, with 6 of these stations being found on the Superterp, a small island, 320 m in diameter. The stations in the Superterp make up the shortest baselines in LOFAR's array. The antennas in the core station are distributed to give optimal \textit{uv} coverage. There are 14 stations, outside of the core, that are spaced out logarithmically, with the outer most station being at a radius of 90 km. These 14 stations are denoted as the `remote stations'\footnote{http://old.astron.nl/radio-observatory/astronomers/users/technical-information/lofar-array-configuration/lofar-array-conf} \citep{LOFAROVERVIEW}.

The observation data used here is from the LOFAR EoR project, concentrated on the North Celestial Pole (NCP) window \citep{WSRT_NCP,Yatawatta}. The data is an accumulation of $\sim$140 hours of observing time, coming from 10 nights worth of observations (observing nights ID: L80847, L86762, L90490, L196421, L205861, L246297, L246309, L253987, L254116, L254865) \citep{MertensUpper2020}. The data has undergone both direction-independent and -dependent calibration as well as RFI flagging \citep{SAGECAL08,SAGECAL10,RFI_FLAG}. A more rigorous description of the data processing pipeline for this dataset can be found in \citet{UpperLim_HBA} and \citet{MertensUpper2020}. This data set has a total bandwidth $\Delta \nu$ = 11.52 MHz, with frequency range 134.57 MHz $\leq \nu \leq$ 146.09 MHz. The frequency resolution of the data is 195.3 kHz, however several sub-bands were flagged resulting in the current 37 sub-band dataset.

\subsection{Simulated Data}
\subsubsection{21cm EoR Signal}
\label{eor_sim}
\begin{figure}
\centering
\includegraphics[width=1\linewidth]{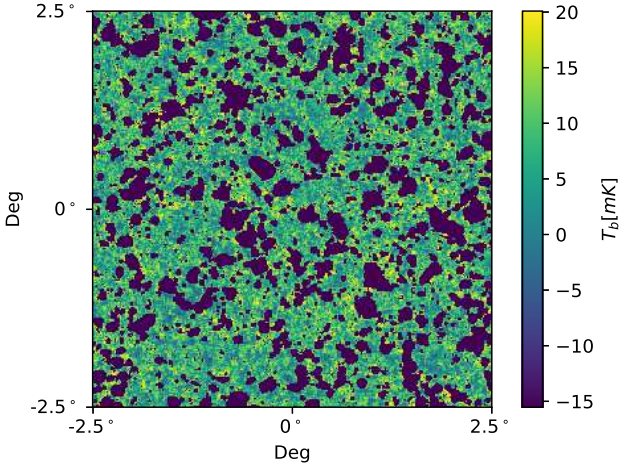}
\caption{A frequency slice at 134 MHz (z $\sim$ 9.6) of the cosmological signal realisation produced by {\fontfamily{cmtt}\selectfont
21CMFAST}, with a neutral fraction x$_{\rm{HI}}$ = 0.78.}
\label{fig:21cmFastSlice}
\end{figure}

\begin{figure}
\centering
\includegraphics[width=1\linewidth]{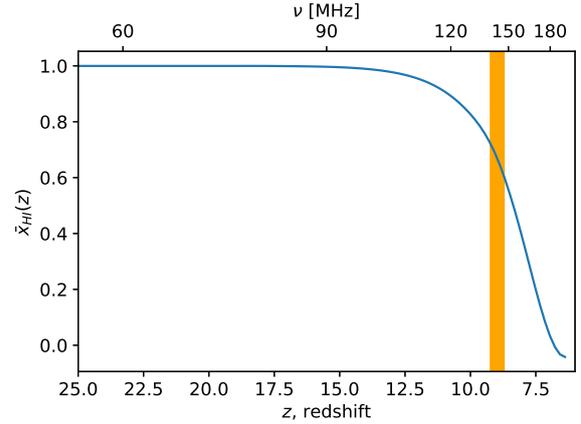}
\caption{The neutral fraction evolution, with redshift, of the cosmological simulation produced by {\fontfamily{cmtt}\selectfont
21CMFAST}. The shaded region highlights the main frequency region of interest: 134.5 MHz $\leq \nu \leq$ 146 MHz.}
\label{fig:neutralx}
\end{figure}

The 21cm signal is simulated using {\fontfamily{cmtt}\selectfont
21CMFAST} \citep{21cmFast}. This is a semi-analytical tool  which approximates  physical processes, rather than simulating via hydrodynamic simulations, making it computationally much less expensive. The results, the evolving brightness temperature T$_b$, produced by {\fontfamily{cmtt}\selectfont
21CMFAST} agree with recent hydrodynamical simulations \citep{Hutter}. The simulation used {\fontfamily{cmtt}\selectfont
21CMFAST}'s default Planck cosmology: ($\sigma_8$,$h$,$\Omega_m$,$\Omega_{\Lambda}$,n) = (0.81,0.68,0.31,0.69,0.97). The simulation was initialised at $z$ = 25, with a box size of 1000 Mpc, on a 2048$^3$ grid with subsequent evolution of the density and ionization being performed on a lower resolution 512$^3$ grid - with this grid being the final T$_b$ box size. We have assumed saturation of the spin temperature T$_s$: T$_s$ $\ggg$ T$_{\rm{CMB}}$, which is valid during the redshifts of the observation; we have set the UV photons per stellar baryon to be 0.3. The output of {\fontfamily{cmtt}\selectfont
21CMFAST} is the $\delta T_b$ field in the redshift range 6$\leq z \leq$25, with these redshift slices being made into a lightcone computed with an observing angle of 5$^{\circ}$ - to match that of LOFAR. Fig. \ref{fig:21cmFastSlice} shows an example slice and the evolution of neutral hydrogen for this realisation is shown in Fig. \ref{fig:neutralx}.

With the known bandwidth of the LOFAR 10 nights data cube, 11.52 MHz, and sub-band width, 195.3 kHz, we simulate the cosmological signal via {\fontfamily{cmtt}\selectfont
21CMFAST} between 134.57 MHz and 146.09 MHz, and then inject this cosmological signal into the data.  


\subsubsection{Foregrounds}
\label{Foregrounds}
\begin{figure}
\centering
\includegraphics[width=1\linewidth]{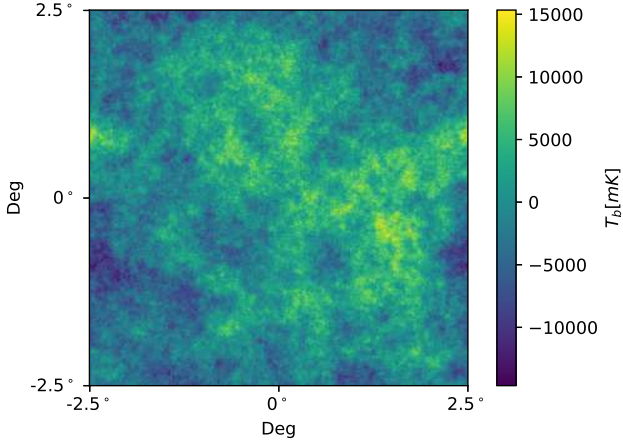}
\caption{A frequency slice at 134 MHz of the foreground simulation produced by the lightcone outlined in \citet{Jov08}.}
\label{fig:Vibor_Slice}
\end{figure}

The diffuse astrophysical foregrounds were simulated using the methods described in \citet{Jov08}. This treats the foreground emissions, within our FoV at a given frequency, as Gaussian random fields (GRF), modelling the foregrounds' emission as a power law: $T_{b} \propto \nu^{-\beta}$. The foreground contributions are the following: 

\begin{description}
  \item[$\bullet$ \textbf{Galactic diffuse synchrotron emission (GDSE)}:]Galactic synchrotron emission is the dominant foreground component that radio surveys encounter. This emission arises from two sources: The diffuse component arises from the interaction between relativistic free electrons from the interstellar medium (ISM) and the Galactic magnetic field. The second is due to emission from supernovae remnants (SNRs). Together these make up $\sim$ 70$\%$ of the observed foreground emission. The brightness temperature spectral index used to model this was $\beta$ = 2.55 - this value was used as the mean value for a GRF with a deviation of 0.1. 
  \item[$\bullet$ \textbf{Galactic diffuse free-free emission}:] This foreground is better known as bremsstrahlung emission and it only accounts for $~ 1 \%$ of the contaminating foregrounds (both Galactic and extra). Bremsstrahlung emission is due to electrons scattering off of other ions. The scattering off ions causes them to decelerate and emit radiation. At the high latitudes observed by radio arrays to detect the 21cm signal, this emission is from diffused ionized gas which is optically thin. When simulated, it has a fixed spectral index of 2.15.
  \item[$\bullet$\textbf{ Extragalactic foregrounds}:] Extragalactic foregrounds account for 27$\%$ of the foregrounds that contaminate our signal. These fall into two categories: 1)  Radio galaxies, observed as point sources in the sky and the source of their observed radio signal is synchrotron emission. 2) Radio clusters, the largest virialised systems in the Universe with the emission from clusters coming from their galaxies and from the intercluster medium (ICM). Radio sources of galaxy clusters have two classifications: radio halos and radio relics. Radio halos have a regular morphology with a low surface brightness, typically centred on and permeating the cluster volume. Radio relics are similar to radio halos, having low surface brightness and extending over large distances ($>$ 1 Mpc). \citet{Jov08} clusters extragalactic foregrounds using a random walk algorithm. 
  
We can see an example slice of the foregrounds in Fig. \ref{fig:Vibor_Slice}. 
\end{description}

The simulated data have a field of view (FoV) of 5$^{\circ}$x 5$^{\circ}$, with a frequency resolution of 500 kHz and pixel resolution of 35 arcseconds.
Henceforth, we will refer to the data, a cube with bandwidth $\Delta\nu$ with a field of view of deg x deg, interchangeably, as a data cube or cube.

\subsubsection{Simulated observations}

The next step from simulating a realisation of reionization is to apply observational effects onto the simulated lightcone. We do this using the OSKAR visibility simulation package\footnote{https://github.com/OxfordSKA/OSKAR}. OSKAR produces simulated visibility data from radio telescopes containing aperture arrays. It does this by taking an input file that contains antenna placement and (permitted) beam positions. Taking these, and a sky model provided, it produces dual-polarisation signals for each antenna provided before using these to beam form and hence simulate an observation. These observational effects are applied to the data provided, in our case the simulated data, to produce a measurement set. The measurement set\footnote{More information on measurement sets can be found: https://casa.nrao.edu/Memos/229.html.} output contains visibility data, produced by OSKAR simulating an observation of our provided data. With the measurement set in hand, we next move on to producing images from these visibilities, and using WSClean \citep{WSClean}. We use OSKAR to simulate a 120 hour observation with a data cube with both the simulated foregrounds and cosmological signal. To produce an image, WSClean takes the provided visibilities and grids them onto different $w$-layers. WSClean attempts to solve the complex visibility function,
\begin{equation}\label{eq:vis_equation}
\begin{split}
V(u,v,w)= &\int\int\frac{A(l,m)I(l,m)}{\sqrt{1-l^2-m^2}} \times\\ 
&\exp  \bigg\{2\pi i \left[ul + vm + w\left(\sqrt{1-l^2-m^2}-1\right)\right]  \bigg\}  dldm
\end{split}
\end{equation}
where $u, v, w$ are the baseline coordinates in the coordinate system of the array, $A(l,m)$ is the primary-beam function, $I(l,m)$ is the sky function and \textit{l,m} are cosine sky coordinates. WSClean solves this, to find $I(l,m)$ by discretising $w$ such that its integral becomes a summation. Once WSClean solves this to produce an image, it outputs it as a Flexible Image Transport System (FITS) file. We have used natural weighting, within WSClean. The visibilities produced by OSKAR are in Janskys, with WSClean converting the data to Jansky per beam. We then convert to Kelvin, using the beam size for each frequency slice of the data cube. As we have used natural weighting, one must account for the PSF; see section 3.2.1 in \citet{MertensUpper2020}. We also note that the field of view of this image cube has been reduced to 4$^{\circ}$x 4$^{\circ}$, such that it is inside the primary beam.

\section{Results}

In this section we show the results of applying the foreground removal techniques to the aforementioned data. We attempt to replicate the recent upper limit in \citet{MertensUpper2020}, and then look at the application of the techniques to simulated data that replicates the 10 nights of LOFAR data. 

The power spectrum, on a wavenumber \textit{k}, is defined as, 
\begin{center}
\begin{equation}\label{eq:power_spectrum}
P(k) = \rm{V_c}|\tilde{T}(\textit{k})|^2,
\end{equation}
\end{center}
where $\tilde{T}(k)$ is the Fourier transformed brightness temperature and $\rm{V_c}$ is the observed comoving cosmological volume, with the primary beam; spatial tapering function, and frequency tapering function delimited. We then average equation \eqref{eq:power_spectrum} in spherical shells and define the spherically averaged dimensionless power spectrum as,
\begin{center}
\begin{equation}\label{eq:21dimlessPS}
\Delta^2(k) = \frac{k^3}{2\pi^2}P(k).
\end{equation}
\end{center}
When referring to the power spectrum, we are referring to equation \eqref{eq:21dimlessPS}. 

We can use the Stokes I and Stokes V from the  dataset's observation to estimate noise. The Noise estimates are made by differencing two frequency slices in the image cubes. As these slices are separated by a small frequency interval, 195.3 kHz, both the foregrounds and 21cm should be near identical between the two - leaving only the noise. The LOFAR point spread function changes by 0.1 percent over 0.2 MHz \citep{Noise_est}. Thus, this difference should be dominated by thermal noise, more notably after foreground removal. Once we have removed our foreground model from the data, we are left with the residual $\Delta^2_I$. We can now subtract the spherically-averaged noise power spectrum  $\Delta^2_N$, also known as noise bias removal, from the residual:
\begin{center}
\begin{equation}\label{eq:NoiseBiasRemoved}
\Delta^2_{21} = \Delta^2_I - \Delta^2_N.
\end{equation}
\end{center}

With the associated error,
\begin{center}
\begin{equation}\label{eq:NoiseBiasRemovedError}
\Delta^2_{21,err} = \sqrt{\Big(\Delta^2_{I,err}\Big)^2 + \Big(\Delta^2_{N,err}\Big)^2}.
\end{equation}
\end{center}

Unless stated otherwise, all power spectra have the noise bias removed. 

\label{result}
\subsection{LOFAR 10 Nights}
\label{result:LOFAR}
\begin{figure}
\centering
\includegraphics[width=1\linewidth]{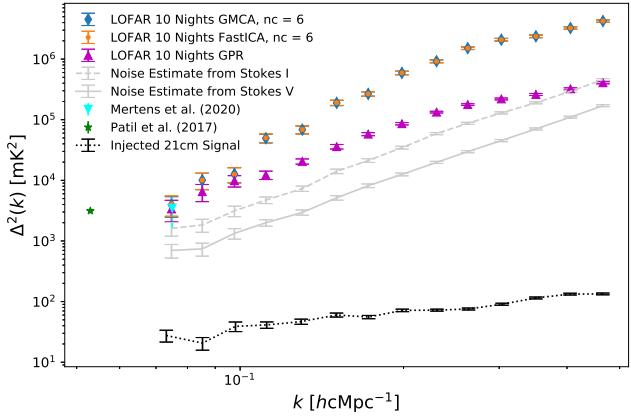}
\caption{Comparison of the noise bias removed residual power spectra produced by the removal techniques: FastICA, GMCA, and GPR. We have also included the power spectrum of the injected cosmological signal and the noise estimates produced by Stokes I and Stokes V. One can see the comparison to recent upper limits produced by LOFAR HBA: \citet{UpperLim_HBA} and \citet{MertensUpper2020}. Within figures involving the BSS techniques, we shall refer to the number of components used to model the foregrounds as `nc'.}
\label{fig:C007010_PS}
\end{figure}

We start with the 10 nights of data (141 hours) from the LOFAR EoR experiment, whose upper limit from this data can be seen in \citet{MertensUpper2020}. In our analysis we compare FastICA, GMCA, and GPR . For FastICA and GMCA, we modelled the foregrounds with a number of different components ranging from 1 to 37; the maximum number of components is set by the dimensionality of the dataset - the number of frequency channels. Here, we present only the result from the best performing number of components, which we find to be 6. We see signal suppression at component numbers higher then this. This is in line with the component number used in \citet{UpperLim_HBA}'s upper limit. For GPR, to model kernels of the data, we apply the reasoning within \citet{MertensUpper2020}: The astrophysical foregrounds, dubbed \textit{intrinsic foregrounds}, are expected to be smooth, with respect to frequency. Thus their frequency coherence-scale will be long, and in this case we have given GPR a uniform prior for the coherency range, this is between between 10-100 MHz, from which to optimise, with a Matern kernel with $\eta_{mix}$ = 5/2. The Bayesian optimisation finds the optimised coherency length for the intrinsic foregrounds to be 46.23 MHz. The second `foreground', comes from instrumentation effects. We have given the  frequency coherence-scale, used for a Matern kernel with $\eta_{mix}$ = 3/2, a range 1-10 MHz. The Bayesian optimisation finds the optimised coherency length for this instrumental foreground to be 2.64 MHz. For the simulated 21cm signal, we use an exponential kernel, with a coherency length range between 0.1-1.2 MHz. The Bayesian optimisation finds the optimised coherency length for the 21cm signal to be 0.73 MHz.

We can see the results of the foreground removal techniques application on the 10 nights of LOFAR data in Fig. \ref{fig:C007010_PS}, where we have removed the noise bias from the results of the removal techniques. Tables \ref{tab:GMCA_C007010_Table}, \ref{tab:FastICA_C007010_Table}, and \ref{tab:GPR_C007010_Table} show the power spectrum results at a given $k$-bin, as well as the 2$\sigma$ upper limit ($\Delta^2_{21,\rm{2\sigma}}$), for GMCA, FastICA, and GPR. We see for the smaller $k$ bins, $k$ <  \textit{h}cMpc$^{-1}$, the three removal methods perform similarly. In fact, we recover the recent upper limit presented in \citet{MertensUpper2020}, with our use of GPR replicating the upper limit. The BSS techniques of FastICA and GMCA reproduce the upper limit at the same \textit{k}-scale, \textit{k} = 0.075 \textit{h}cMpc$^{-1}$, with a 2$\sigma$ value of $\Delta^2_{21} <$ (73)$^2$ mK$^2$. This is interesting, as neither of the two assume anything about the data a-priori, whilst GPR requires additional information about the data it is fitting. As we move to \textit{k}-scales larger than  \textit{h}cMpc$^{-1}$, both FastICA and GMCA begin to deviate away from GPR, which consistently stays close to the noise limit (shown in grey). In fact, for each adjacent $k$-bin, at \textit{k}-scales larger than  \textit{h}cMpc$^{-1}$, FastICA and GMCA produce progressively larger powers. FastICA and GMCA perform near identically, with their power over the range of $k$-scales, shown in Fig. \ref{fig:C007010_PS}, being near-identical.

\begin{table}
\centering
\begin{tabular}{|llll|}
\hline
$k$ & $\Delta^2_{21}$ & $\Delta^2_{21,err}$ & $\Delta^2_{21,\rm{2\sigma}}$\\
(\textit{h}cMpc$^{-1}$) & (mK$^2$) & (mK$^2$) & (mK$^2$)\\
\hline
0.075 & (58.30)$^2$ & (31.01)$^2$ & (72.95)$^2$ \\
0.085 & (100.22)$^2$ & (54.93)$^2$ & (126.80)$^2$ \\
0.098 & (112.64)$^2$ & (59.88)$^2$ & (140.92)$^2$ \\
0.112 & (222.53)$^2$ & (89.71)$^2$ & (256.16)$^2$ \\
0.130 & (262.13)$^2$ & (101.36)$^2$ & (298.77)$^2$ \\
0.151 & (435.36)$^2$ & (137.00)$^2$ & (476.52)$^2$ \\
0.172 & (516.08)$^2$ & (150.06)$^2$ & (558.01)$^2$ \\
0.199 & (769.19)$^2$ & (196.27)$^2$ & (817.74)$^2$ \\
0.230 & (960.30)$^2$ & (221.44)$^2$ & (1010.07)$^2$ \\
0.263 & (1230.90)$^2$ & (272.06)$^2$ & (1289.63)$^2$ \\
0.303 & (1444.84)$^2$ & (326.61)$^2$ & (1516.87)$^2$ \\
0.351 & (1541.69)$^2$ & (341.94)$^2$ & (1615.75)$^2$ \\
0.406 & (1809.92)$^2$ & (388.27)$^2$ & (1884.69)$^2$ \\
0.466 & (2058.63)$^2$ & (445.30)$^2$ & (2152.80)$^2$\\
\hline
\end{tabular}
\caption{Showing the $\Delta^2_{21}$ upper limit at 2$\sigma$ ($\Delta^2_{21,\rm{2\sigma}}$), and error ($\Delta^2_{21,err}$), at a given $k$-bin for GMCA.}
\label{tab:GMCA_C007010_Table}
\end{table}

\begin{table}
\centering
\begin{tabular}{|llll|}
\hline
$k$ & $\Delta^2_{21}$ & $\Delta^2_{21,err}$ & $\Delta^2_{21,\rm{2\sigma}}$\\
(\textit{h}cMpc$^{-1}$) & (mK$^2$) & (mK$^2$) & (mK$^2$)\\
\hline
0.075 & (58.24)$^2$ & (31.10)$^2$ & (72.98)$^2$ \\
0.085 & (100.32)$^2$ & (54.98)$^2$ & (126.92)$^2$ \\
0.098 & (112.00)$^2$ & (59.57)$^2$ & (140.14)$^2$ \\
0.112 & (224.05)$^2$ & (90.30)$^2$ & (257.89)$^2$ \\
0.130 & (262.41)$^2$ & (101.46)$^2$ & (299.07)$^2$ \\
0.151 & (436.77)$^2$ & (137.43)$^2$ & (478.06)$^2$ \\
0.172 & (516.43)$^2$ & (150.16)$^2$ & (558.39)$^2$ \\
0.199 & (769.99)$^2$ & (196.21)$^2$ & (817.52)$^2$ \\
0.230 & (959.09)$^2$ & (221.16)$^2$ & (1008.80)$^2$ \\
0.263 & (1228.46)$^2$ & (271.53)$^2$ & (1287.08)$^2$ \\
0.303 & (1443.43)$^2$ & (326.29)$^2$ & (1514.40)$^2$ \\
0.351 & (1540.83)$^2$ & (341.75)$^2$ & (1614.85)$^2$ \\
0.406 & (1803.51)$^2$ & (388.39)$^2$ & (1885.29)$^2$ \\
0.466 & (2059.14)$^2$ & (445.41)$^2$ & (2153.33)$^2$\\
\hline
\end{tabular}
\caption{Showing the $\Delta^2_{21}$ upper limit at 2$\sigma$ ($\Delta^2_{21,\rm{2\sigma}}$), and error ($\Delta^2_{21,err}$), at a given $k$-bin for FastICA.}
\label{tab:FastICA_C007010_Table}
\end{table}

\begin{table}
\centering
\begin{tabular}{|llll|}
\hline
$k$ & $\Delta^2_{21}$ & $\Delta^2_{21,err}$ & $\Delta^2_{21,\rm{2\sigma}}$\\
(\textit{h}cMpc$^{-1}$) & (mK$^2$) & (mK$^2$) & (mK$^2$)\\
\hline
0.075 & (58.06)$^2$ & (31.20)$^2$ & (72.92)$^2$ \\
0.085 & (80.61)$^2$ & (33.45)$^2$ & (93.47)$^2$ \\
0.098 & (99.22)$^2$ & (35.03)$^2$ & (110.90)$^2$ \\
0.112 & (110.65)$^2$ & (44.04)$^2$ & (126.97)$^2$ \\
0.130 & (143.23)$^2$ & (44.97)$^2$ & (156.72)$^2$ \\
0.151 & (189.54)$^2$ & (55.69)$^2$ & (205.24)$^2$ \\
0.172 & (240.49)$^2$ & (57.87)$^2$ & (254.03)$^2$ \\
0.199 & (293.08)$^2$ & (61.88)$^2$ & (305.86)$^2$ \\
0.230 & (365.19)$^2$ & (65.23)$^2$ & (376.66)$^2$ \\
0.263 & (421.48)$^2$ & (75.85)$^2$ & (434.91)$^2$ \\
0.303 & (469.14)$^2$ & (88.26)$^2$ & (485.46)$^2$ \\
0.351 & (510.65)$^2$ & (99.14)$^2$ & (529.55)$^2$ \\
0.406 & (568.37)$^2$ & (112.48)$^2$ & (590.21)$^2$ \\
0.466 & (638.14)$^2$ & (129.66)$^2$ & (663.73)$^2$\\
\hline
\end{tabular}
\caption{Showing the $\Delta^2_{21}$ upper limit at 2$\sigma$ ($\Delta^2_{21,\rm{2\sigma}}$), and error ($\Delta^2_{21,err}$), at a given $k$-bin for GPR.}
\label{tab:GPR_C007010_Table}
\end{table}

\subsection{Simulated Observation}
\label{result:Sim}

From the results in Fig. \ref{fig:C007010_PS}, we see that both FastICA and GMCA begin to diverge from the result of GPR at \textit{k}-scales larger than  \textit{h}cMpc$^{-1}$. To better understand why, we further explore the performance of both GMCA and FastICA, with the performance scales $k$ >  \textit{h}cMpc$^{-1}$ being the motif throughout the analysis. 

Within our plots we have a few graphical convention: Any \textit{k}-bins where the power of the residuals, before noise bias removal, is below that of the noise are omitted. This leads to an unphysical negative power for the residual once the noise bias has been removed; any error bars extending to below the x axis in log space are shown with equal-sized arrows to indicate this.

\subsubsection{Simulated: Clean}
\begin{figure}
\centering
\includegraphics[width=1\linewidth]{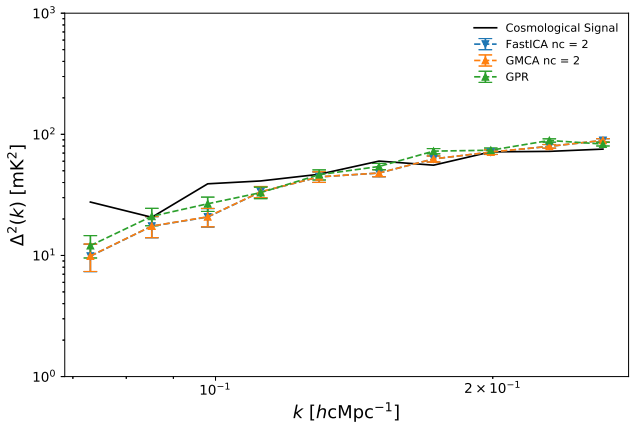}
\caption{Comparison of FastICA, GMCA, and GPR residual power spectra as they recover the cosmological signal injected into simulated foregrounds (outlined in \citet{Jov08}) - recovery performed over a $\Delta \nu$ = 30 MHz bandwidth, 120 MHz $\leq \nu \leq$ 150 MHz. We model the foregrounds with two components for FastICA and GMCA, determined by the method outlined in Section \ref{result:Sim}. Due to their similar performance, the FastICA result has been overlaid by GMCA. }
\label{fig:Sim_Comp}
\end{figure}

We first start with how the foreground removal techniques perform on simulated data that has not gone through OSKAR nor had noise added; this data has both the cosmological signal and foregrounds, only. We can see the result of the power spectrum recovery in Fig. \ref{fig:Sim_Comp}.

To find the best number of components to model the foregrounds with the BSS techniques, we applied the following technique for the goodness of fit: Taking the squared difference between the residual of our BSS techniques, being the recovered cosmological signal, and the cosmological signal we wish to recover. We sum over each pixel of the data cube's squared difference, and normalise this difference by the variance of the cosmological signal at the frequency the pixel lies in. If the residual perfectly matches the cosmological signal, the returned value is zero. The motivation for normalising by the variance allows us to see if the squared error in the residual is below that of signals variance, giving a value below 1, and hence the cosmological signal is still detectable. 

We find that modelling the foregrounds, within FastICA and GMCA, is best with 2 components. For GPR, we use the same covariance matrix described in Section \ref{result:LOFAR} but without the mode-mixing kernel. All three of the techniques perform near identically, and do recover the 21cm power spectrum fairly well, with their aforementioned technique for goodness of fit returning a value below 0.1.
\subsubsection{Simulated: Observation Comparison}

We now look to see how well FastICA and GMCA perform on simulated observations via OSKAR. We compare three simulated observations with noise included:
\begin{enumerate}
  \item 120 hour observation.
  \item 120 hour observation but we have removed the primary beam from the observation.
  \item 120 hour observation, but we have used the resolution from the lowest frequency in the observation’s point source function (PSF). For the purposes of this paper it is the PSF at 134.5 MHz; the primary beam is also included.
\end{enumerate}

\begin{figure}
     \centering
     \begin{subfigure}[b]{0.5\textwidth}
         \centering
         \includegraphics[width=\textwidth]{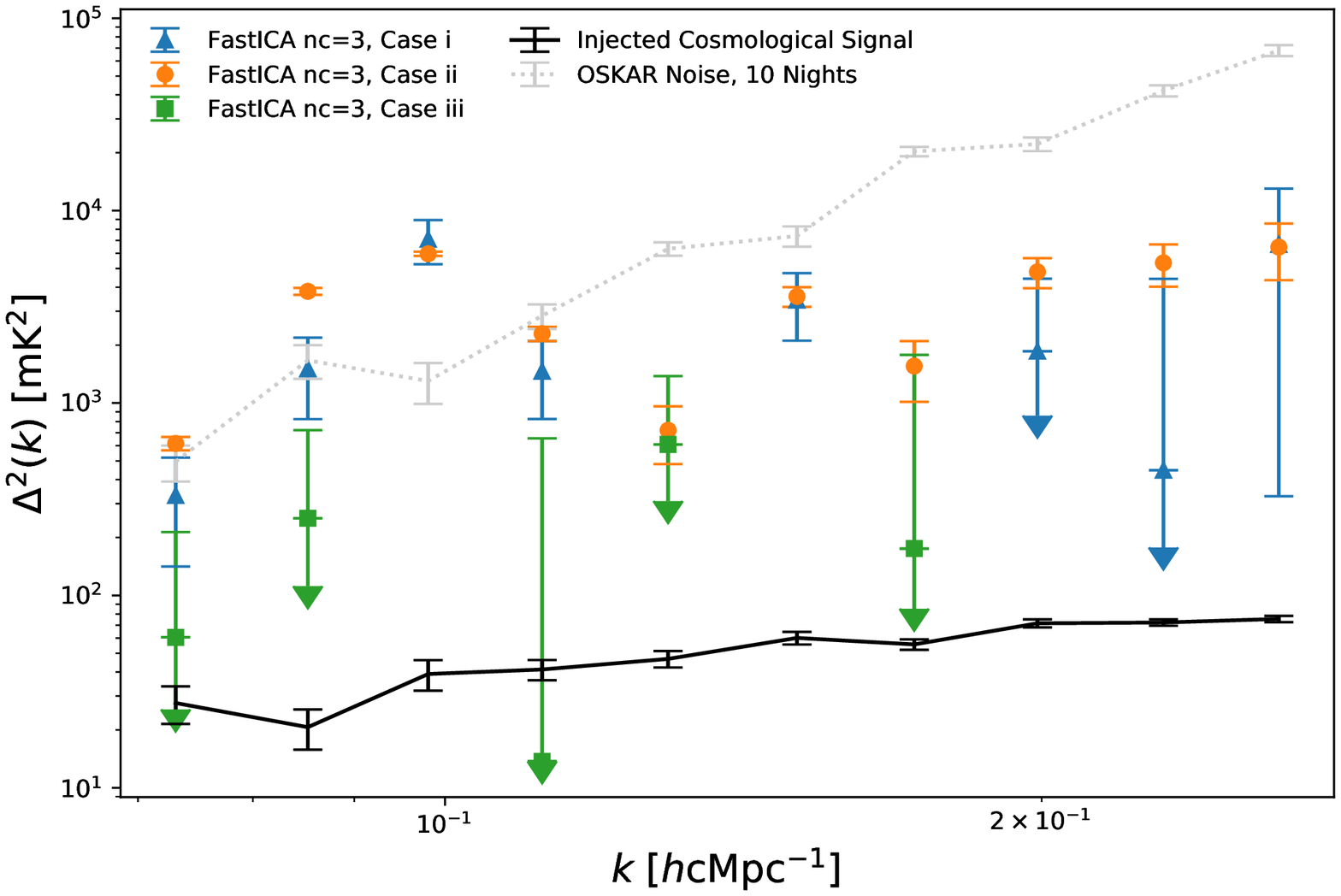}
         \caption{FastICA}
         \label{fig: Case FastICA}
     \end{subfigure}
     \hfill
     \begin{subfigure}[b]{0.5\textwidth}
         \centering
         \includegraphics[width=\textwidth]{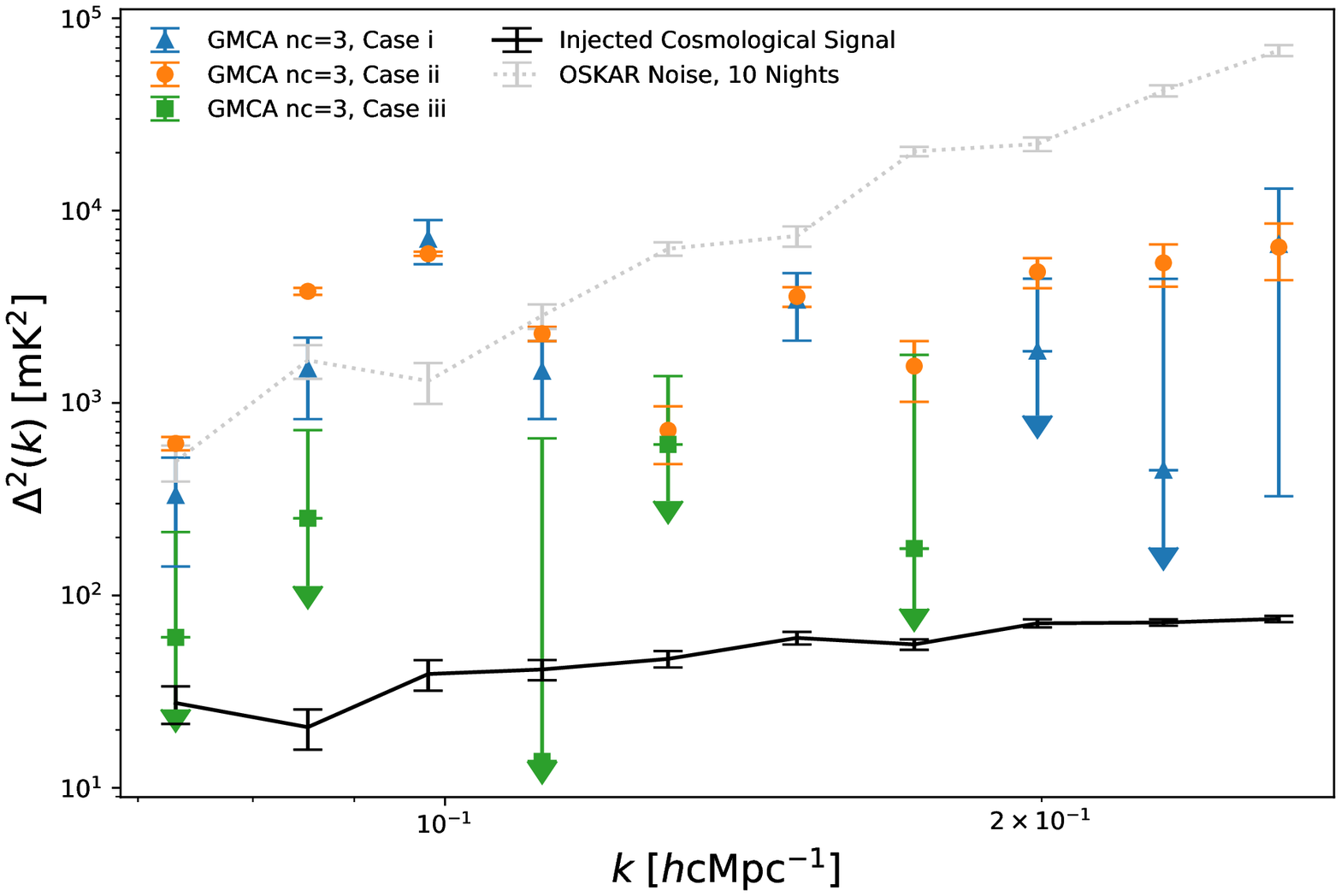}
         \caption{GMCA}
         \label{fig:Case GMCA}
     \end{subfigure}
     \hfill
     \begin{subfigure}[b]{0.5\textwidth}
         \centering
         \includegraphics[width=\textwidth]{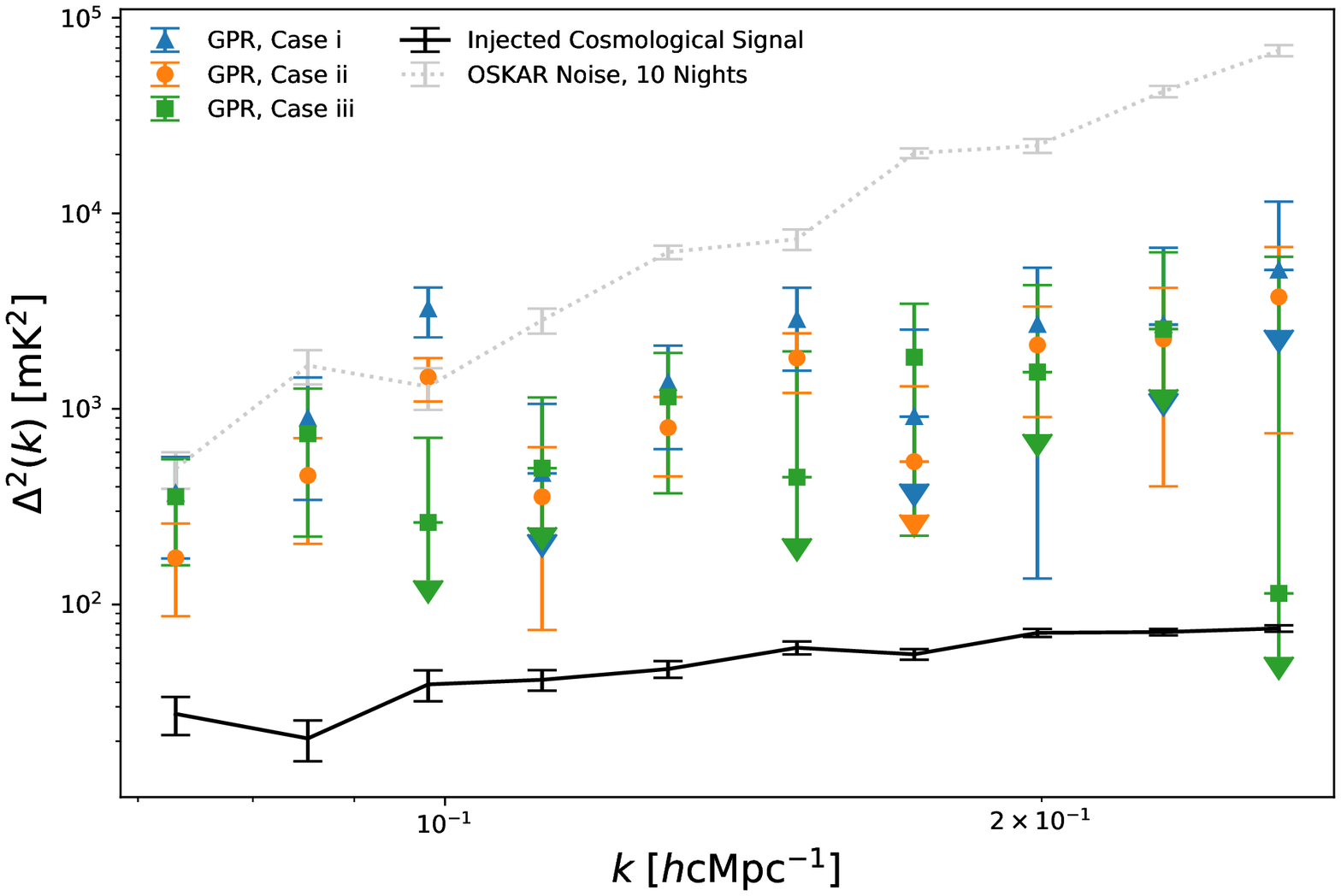}
          \caption{GPR}
         \label{fig:Case GPR}
     \end{subfigure}
    \caption{Comparison of FastICA, GMCA, and GPR in recovering a cosmological signal injected into simulated foregrounds (outlined in \citet{Jov08}), with OSKAR being used to simulate observations. We then compare three cases: \textit{i}) Normal 120 hour observation.  \textit{ii}) 120 hour observation with no primary beam included.  \textit{iii}) 120 hour observation with the PSF of the lowest frequency (134.5 MHz), with the primary beam included. The bandwidth of the cube is $\Delta\nu$ = 11.5 MHz, in the range: 134.5 MHz $\leq \nu \leq$ 146 MHz.}
    \label{fig:Sim_Comp_Obs}
\end{figure}

We see the results of this in Fig. \ref{fig:Sim_Comp_Obs}. The reasoning for case \textit{ii} is to see whether the primary beam effects are the limiting factor for both FastICA and GMCA (compared to GPR), and case \textit{iii} is to see if the chromaticity of the instrument is a limiting factor for FastICA and GMCA. As mentioned in Section \ref{result:LOFAR}, GPR takes kernels for the various components in the data, one of which was \textit{\textbf{K$_{\rm{fg}}$}}, which is the kernel for the instrumental effects - namely, mode-mixing. Whilst GPR accounts for this, FastICA and GMCA, being BSS techniques, assume nothing of the data a-priori and solely attempt to fit the data. As it is ignorant to such effects, they may provide a hindrance to the recovery capabilities of the two techniques. For case \textit{iii}, we have removed this mode-mixing covariance from GPR's modelling as we no longer have the chromaticity of the instrument.

We only show the `best' (i.e. closest to the noise limit) results. For the BSS techniques, the number of components used to define the foregrounds ranged from 1 to 37; with the number of components being limited by the dimensionality of the dataset - which, here, is set by the number of frequency channels. We choose the number of components to be 3, as this is what we find to be best for case \textit{i}. We perform the analysis of cases \textit{ii} and \textit{iii} with 3 components to compare how the results change. From Fig. \ref{fig:Sim_Comp_Obs}, it is clear to see that case \textit{i}, which represents a normal observation, has a better result for FastICA and GMCA than that in Fig. \ref{fig:C007010_PS}. This result is due to a common downfall of simulations, as they do no reproduce residuals that arise from a normal LOFAR pipeline, such as direction-dependent and -independent calibration errors or RFI flagging. \citet{2019:OffringaMertensKoopmans} showed that RFI flagging can lead to a \textit{flagging excess power} - which was found to be mitigated by the use of a Gaussian-weighted interpolation scheme for the flagged samples along with a unitary weighting during averaging. They also showed that that GPR is able to model some of this excess at low \textit{k}-scales, and so it is a reasonable question to ask if the effect seen at \textit{k}-scales larger than  \textit{h}cMpc$^{-1}$ is due to this. We tested this by taking an extreme example of $\sim \frac{1}{3}$ of the channels flagged, which is the equivalent of the gridding step with flagging. We found that, for the case \textit{i} for GMCA and FastICA, there is excess at scales of $\sim$ 0.2 \textit{h}cMpc$^{-1}$ and larger, progressively taking the residual to the noise limit by the penultimate \textit{k}-bin. We did not see any change in the GPR result, between the flagged test and Fig. \ref{fig:Case GPR}. Seeing as the data used in Fig. \ref{fig:C007010_PS} used the steps proposed by \citet{2019:OffringaMertensKoopmans}, we can be sure it is not this effect causing it. From Figs. \ref{fig: Case FastICA} and \ref{fig:Case GMCA}, we see, as expected, FastICA and GMCA perform near identically. For case \textit{ii}, one should note that we do not delimit the primary beam from the power spectrum (equation 22, \citet{MertensUpper2020}). We see that the results for case \textit{ii} are similar to that of case \textit{i}. This is not surprising as we reduce our FoV from 5$^{\circ}$x 5$^{\circ}$ to 4$^{\circ}$x 4$^{\circ}$ to be within the primary beam, and so the effects of the primary beam are less significant. For case \textit{iii}, the removal of chromaticity effects allow FastICA and GMCA to perform better on the lowest \textit{k}-scales, as mode mixing introduces additional fluctuations into the observed signal - that are no longer present here. However, at larger \textit{k}-scales, we see over-fitting and we see more of the cosmological signal leaked into the foreground model. The removal of the primary beam and chromaticity effects reduces the number of independent components at higher \textit{k}-scales, as such 3 components at higher \textit{k} may be too many, leading to overfitting.

As seen for FastICA and GMCA, case \textit{ii}, unsurprisingly, with no primary beam GPR performs similarly to case \textit{i}. For case \textit{iii}, for which we do not include the mode-mixing kernel, we see that the results are similar to case \textit{i}. We conclude that instrumental effects are not the dominant effect causing the poor recovery of FastICA and GMCA, at scales $k$ >  \textit{h}cMpc$^{-1}$ in Fig. \ref{fig:C007010_PS}. We see instrumental effects yield no worsening of performance on scales $k$ > 0.1 \textit{h}cMpc$^{-1}$. 
With instrumental effects not providing any clarity on the performance of FastICA and GMCA at scales $k$ >  \textit{h}cMpc$^{-1}$, we extend our analysis to look at the further issues, such as limitations due to bandwidth of data and noise.

\subsubsection{Simulated: Not Enough Data}

\begin{figure}
     \centering
     \begin{subfigure}[b]{0.5\textwidth}
         \centering
         \includegraphics[width=\textwidth]{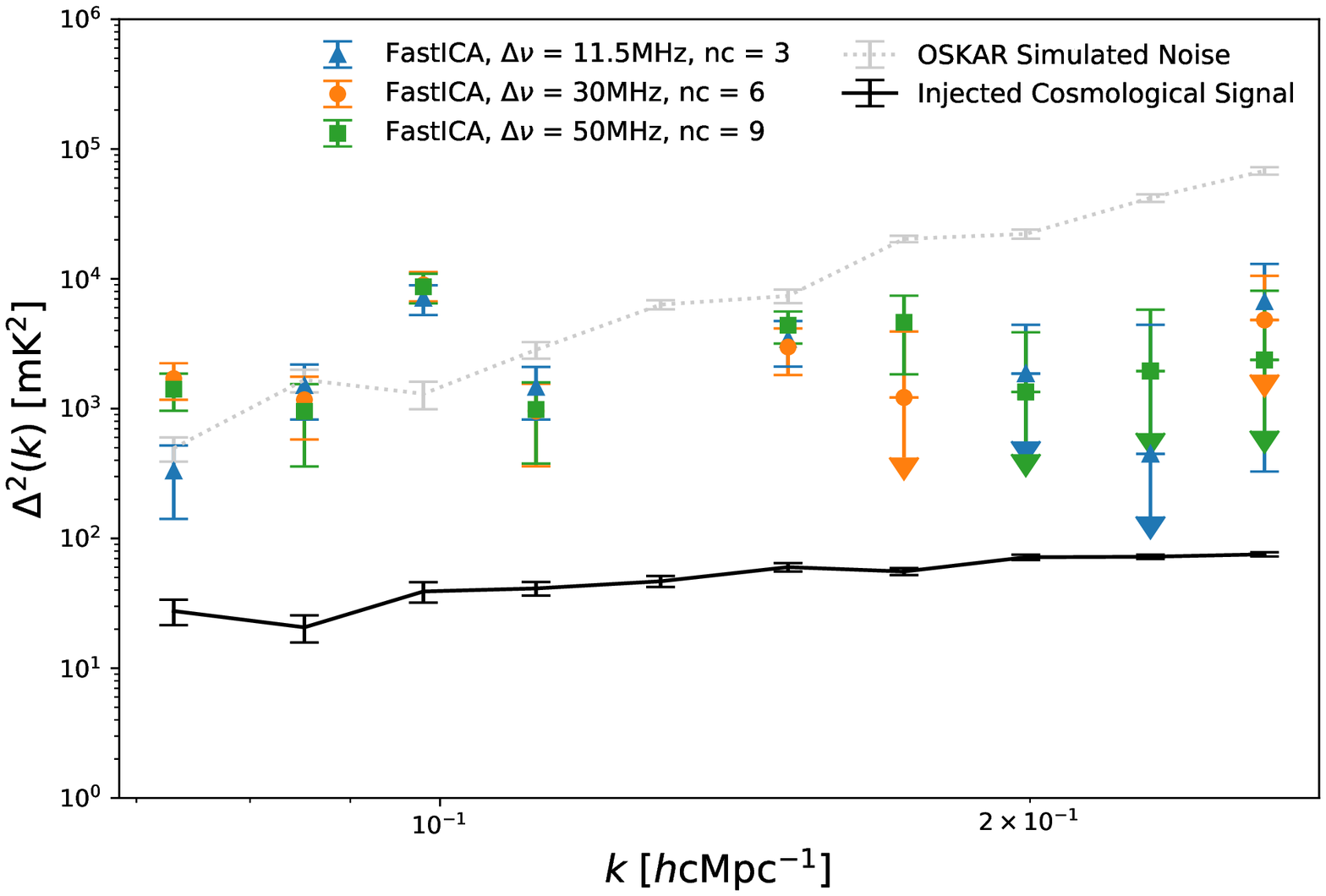}
         \caption{FastICA}
         \label{fig:Delta_nu_comp FastICA}
     \end{subfigure}
     \hfill
     \begin{subfigure}[b]{0.5\textwidth}
         \centering
         \includegraphics[width=\textwidth]{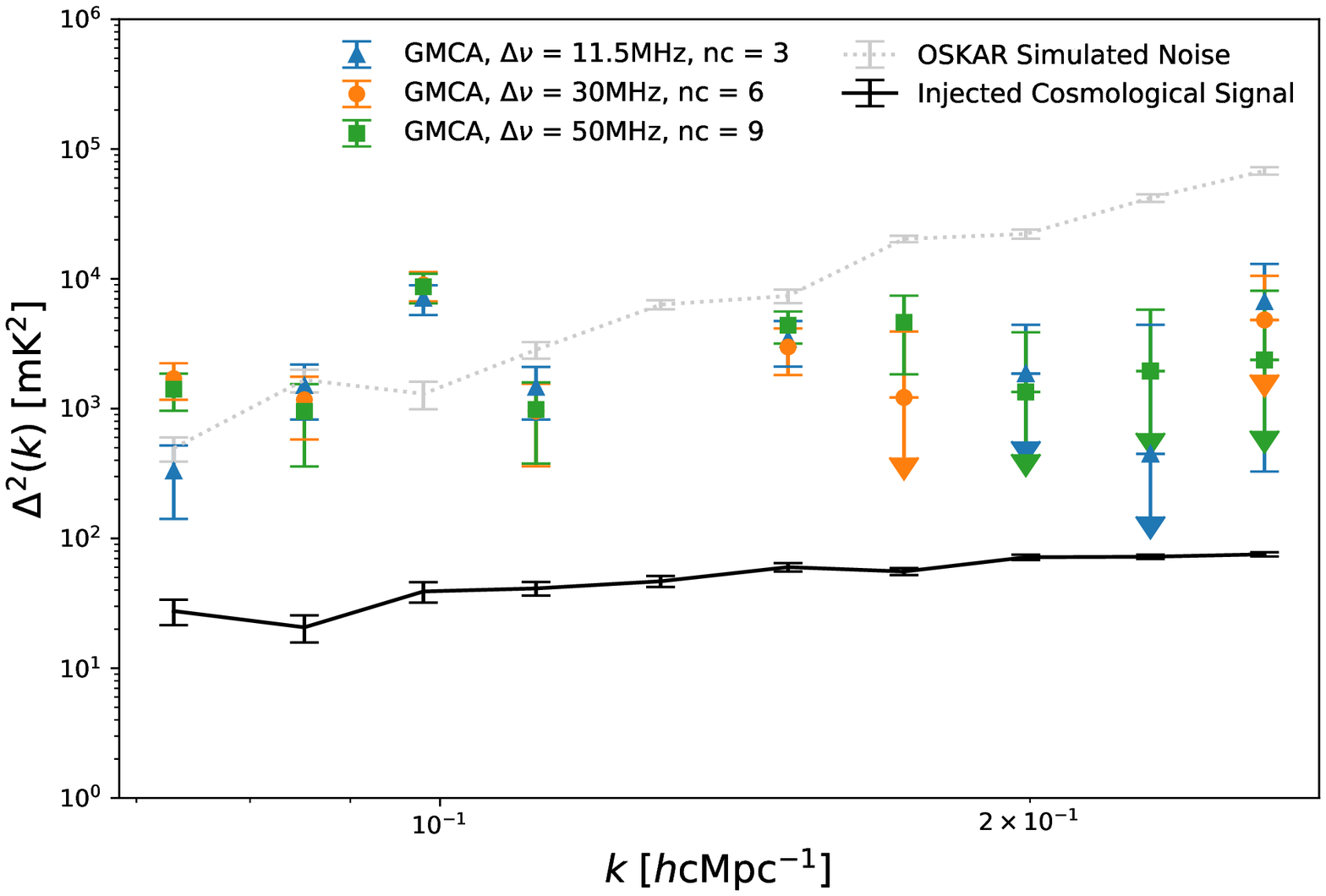}
         \caption{GMCA}
         \label{fig:Delta_nu_comp GMCA}
     \end{subfigure}
     \hfill
     \begin{subfigure}[b]{0.5\textwidth}
         \centering
         \includegraphics[width=\textwidth]{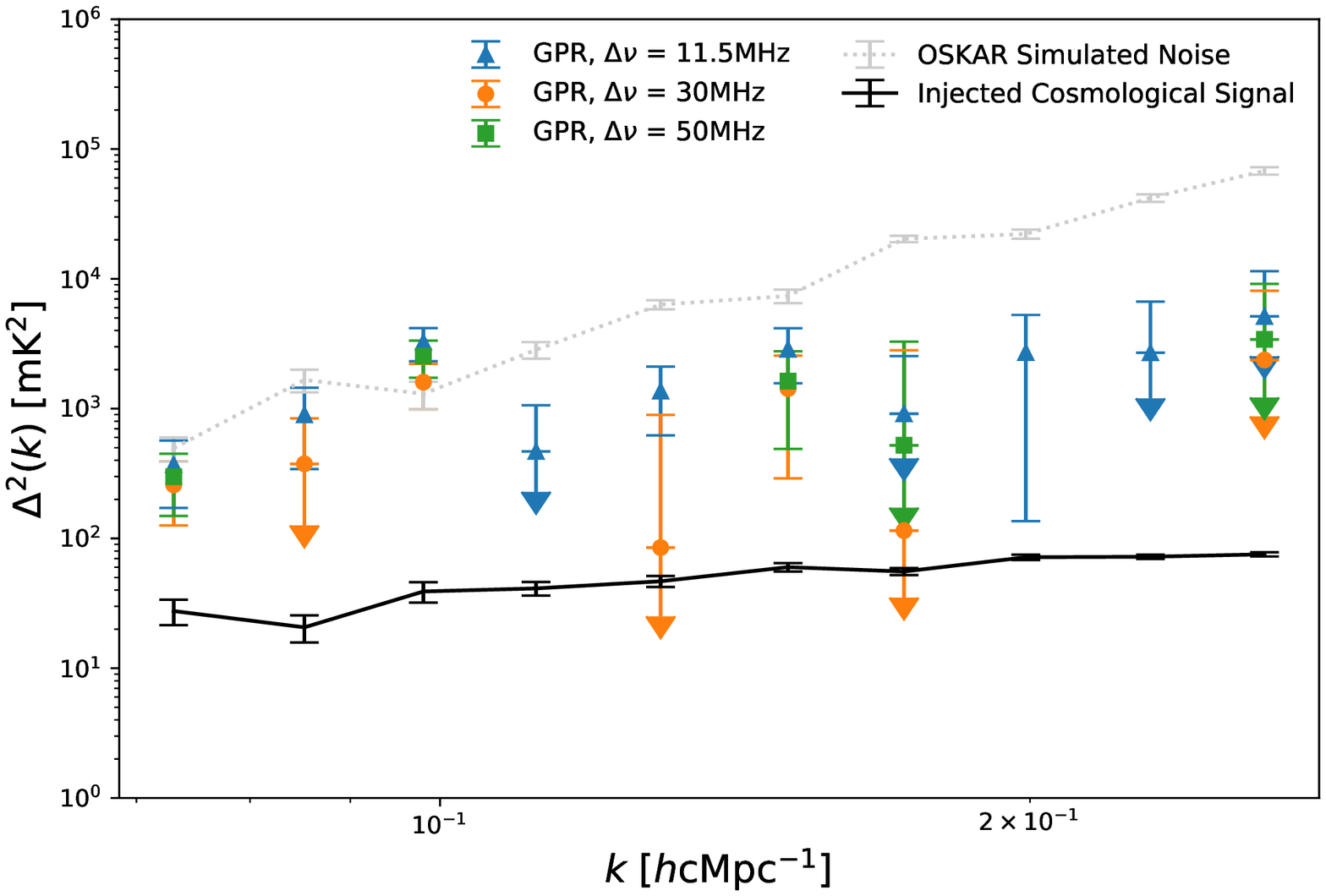}
         \caption{GPR}
         \label{fig:Delta_nu_comp GPR}
     \end{subfigure}
    \caption{Comparison of FastICA, GMCA, and GPR in recovering a cosmological signal injected into simulated foregrounds (outlined in \citet{Jov08}) with OSKAR being used to simulate observations. We compare three cases: The first being the normal $\Delta \nu$ = 11.5 MHz we have been using in out analysis thus far; the second case is a cube of bandwidth $\Delta \nu$ = 30 MHz, and the final case is a cube of bandwidth $\Delta \nu$ = 50 MHz.}
    \label{fig:Delta_nu_comp}
\end{figure}

Currently, the foreground removal techniques have been applied to an image cube of bandwidth $\Delta \nu$ = 11.5 MHz. The more data we give these techniques, the more information they have to fit the foregrounds and, potentially, recover the cosmological signal. We compare three cases: The first being the normal $\Delta \nu$ = 11.5 MHz we have been using in our analysis thus far; the second case is a cube of bandwidth $\Delta \nu$ = 30 MHz, and the final case is a cube of bandwidth $\Delta \nu$ = 50 MHz. We see the results of this in Fig. \ref{fig:Delta_nu_comp}, note that all power spectra are taken in the $\Delta \nu$ = 11.5 MHz bandwidth, 134.5 MHz $\leq \nu \leq$ 146.0 MHz.

From Figs. \ref{fig:Delta_nu_comp FastICA} and \ref{fig:Delta_nu_comp GMCA}, we see there is some improvement with the larger bandwidth, with no clear improvement as we compare the $\Delta\nu$ = 30 MHz and $\Delta\nu$ = 50 MHz for the image cubes for GMCA and FastICA. We choose the number of components such that the cosmological signal is not suppressed, i.e. the cosmological signal doesn't leak into the foreground model. For each of the \textit{k}-bins in the power spectrum, we see that there is no clear improvement of the power spectrum, at scales $k$ >  \textit{h}cMpc$^{-1}$, relative to that if the lower $k$-scales, as we increase the bandwidth. For GPR we see that, overall, the larger bandwidths perform the worse with many \textit{k}-bins in the power spectrum being completely suppressed, Fig. \ref{fig:Delta_nu_comp GPR}. This result with the larger bandwidths is not surprising, as the frequency coherency of the signal evolves with redshift. Using a longer bandwidth would average this evolution, and one would expect a degradation of performance with bandwidth.  

One would expect larger bandwidths to do better as there is more information for GMCA and FastICA to exploit in their decomposition. As aforementioned, GMCA has had great success in its application to CMB data. However, most applications of these BSS techniques have used large bandwidths. In \citet{FASTICA12a} and \citet{GMCA13}, FastICA and GMCA were applied to a dataset of bandwidth $\Delta\nu$ = 85 MHz. In a most recent application of GMCA, \citet{2020:CarucciIrfanBobin}, a dataset of bandwidth $\Delta\nu$ = 400 MHz was used; they also found that having a greater number of channels is better to perform sparsity. So it is likely, out of the realm of the motif of this analysis, that we are not giving the BSS techniques enough data, as even a bandwidth $\Delta\nu$ = 50 MHz seems inadequate. The performance at longer bandwidths will not be affected by the evolution of the signal, such as that which the power spectrum suffers from as we perform it over larger bandwidths \citep{EvolPS1,EvolPS2,2015:GharaDattaChoudhury}. We have seen this when looking at different bandwidth sizes for simulated data that hasn't been applied to OSKAR; as we increased the bandwidth over which removal is performed, the performance of FastICA and GMCA improved. With no clear improvement of the power spectrum, at scales $k$ >  \textit{h}cMpc$^{-1}$ with an increase of bandwidth, we next look at the role of instrumental noise.

\subsubsection{Simulated: Noise Limited}

\begin{figure}
     \centering
     \begin{subfigure}[b]{0.5\textwidth}
         \centering
         \includegraphics[width=\textwidth]{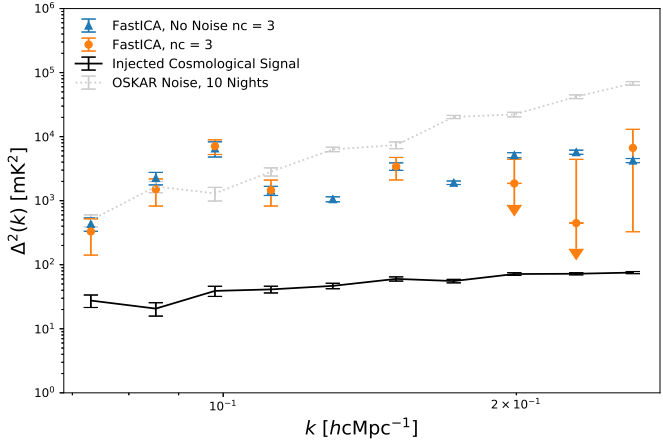}
         \caption{FastICA}
         \label{fig:No Noise FastICA}
     \end{subfigure}
     \hfill
     \begin{subfigure}[b]{0.5\textwidth}
         \centering
         \includegraphics[width=\textwidth]{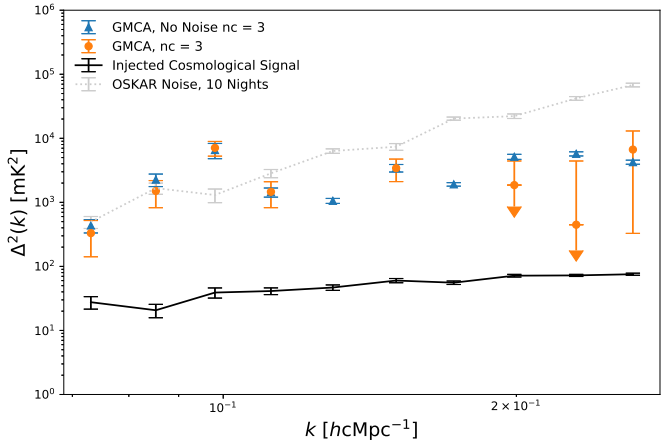}
         \caption{GMCA}
         \label{fig:No Noise GMCA}
     \end{subfigure}
    \caption{Comparing FastICA and GMCA by considering the case where there's no noise. The data has still gone through OSKAR to simulated 10 nights of observation. We show the results using a fiducial value of 3 components to model the foregrounds.}
    \label{fig:No Noise}
\end{figure}

A key difference between the BSS techniques used and GPR, is that the BSS techniques are agnostic to the existence of the cosmological signal in the dataset, whilst GPR not only optimises itself for the inclusion of the 21cm signal but also uses the Stokes V to estimate the noise. This could potentially mean that noise limitations on FastICA and GMCA are greater than GPR, as the former two cast the cosmological signal in the `Gaussian' noise of their model. We now look at the case of no noise. In Fig. \ref{fig:No Noise}, we show  the application of FastICA and GMCA when applied to simulated data containing a no noise. We limit the plot to a fiducial value of 3 components. For FastICA, Fig. \ref{fig:No Noise FastICA}, we see that the no noise case does produce a similar power spectrum to the case with noise. We find that the case with no noise for GMCA, Fig. \ref{fig:No Noise GMCA}, performs the same as the case with noise. However, without the inclusion of noise, they are still orders of magnitude above the injected cosmological signal. Whilst noise is often a limiting factor for recovery, it is not the dominant reason here as to why we see an under-performance on larger \textit{k}-scales, $k$ >  \textit{h}cMpc$^{-1}$. 

An important assumption of the blind source separation technique, as previously implemented on EoR data, is perfect noise avoidance. This is the assumption that we model the noise perfectly, and it does not leak into our foreground model. However, we see that above 4 components noise does begin to leak into the foreground model. This poses a problem, as we can see from Fig. \ref{fig:No Noise}, 3 components is not sufficient to model the foregrounds well enough by the residual - for Fig. \ref{fig:C007010_PS}, 6 components was found to be the optimal number for FastICA and GMCA.

.

\subsection{Blind Source Separation Techniques: Scale Independence}
\label{BSS_Scale_Ind}

\begin{figure}
     \centering
     \begin{subfigure}[b]{0.48\textwidth}
         \centering
         \includegraphics[width=\textwidth]{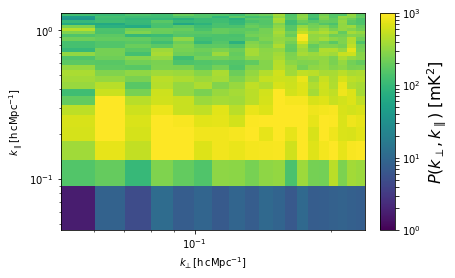}
         \caption{The 2D cylindrical power spectrum of the FastICA foreground removal residual  produced using the LOFAR 10 Nights data, using 6 components to model the foregrounds.}
         \label{fig:2D FastICA}
     \end{subfigure}
     \hfill
     \begin{subfigure}[b]{0.48\textwidth}
         \centering
         
         \includegraphics[width=\textwidth]{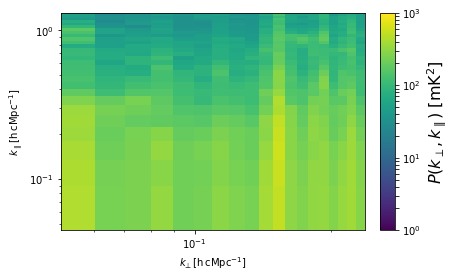}
         \caption{The 2D cylindrical power spectrum of the GPR foreground removal residual produced using the LOFAR 10 Nights data.}
         \label{fig:2D GPR}
     \end{subfigure}
     \hfill
     \begin{subfigure}[b]{0.48\textwidth}
         \centering
         
         \includegraphics[width=\textwidth]{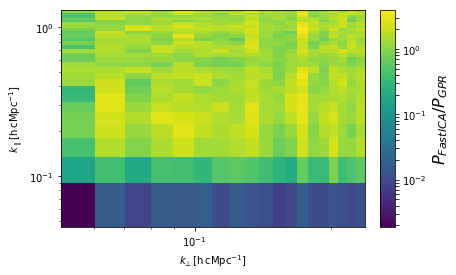}
         \caption{Looking at the 2D cylindrical power spectrum for ratio between FastICA's and GPR's foreground removal residual of the LOFAR 10 Nights Data.}
         \label{fig:2D Ratio}
     \end{subfigure}
    \caption{We compare the dimensionless cylindrical power spectrum of the foreground fitting residuals produced by the application of FastICA, using 6 components to model the foregrounds, and GPR on the 10 nights of LOFAR data.}
    \label{fig:10 Nights 2D}
\end{figure}

Thus far, we have only looked at the performance of the BSS techniques with respect to the data - both its constituents and the amount of data. In Fig. \ref{fig:C007010_PS}, we see that FastICA and GMCA under-perform, with respect to GPR, as we increase the \textit{k}-scale; with each adjacent k-bin producing a higher power than the bin that proceeded it. In Fig. \ref{fig:10 Nights 2D}, we show the foreground removal residuals produced by FastICA and GPR in the 2D cylindrical power spectrum form. As seen throughout this work, FastICA and GMCA perform near-identically; this similar performance was also seen by \citet{2020:CunningtonIrfanCarucci}. As such, for brevity, we only compare FastICA and GPR. The cylindrical power spectrum is the  cylindrically averaged power spectrum and is defined as a function of angular ($k_\perp$) versus line-of-sight ($k_\parallel$) scales as,
\begin{center}
\begin{equation}\label{eq:21cylin}
P(k_\perp,k_\parallel) = \langle P(\textbf{k})_{k_\perp,k_\parallel} \rangle
\end{equation}
\end{center}

In Fig. \ref{fig:2D Ratio}, we see that confined to the lower $k_\parallel$ modes, FastICA gives a comparable removal residual to GPR. However, as we increase $k_\parallel$ and $k_\perp$ we see that FastICA produces a larger power for these modes - in line with what is seen in Fig. \ref{fig:C007010_PS}. At $k_\parallel \geq$ 0.8 \textit{h}cMpc$^{-1}$, we have artefacts introduced by the frequency Fourier transform, which arise from the discontinuities in the frequency sampling (from RFI flagging) causing spectral leakage. Interestingly, at these higher $k_\parallel$, we see more of the spectral leakage in the FastICA residual compared to that of GPR.

\begin{figure}
\centering
\includegraphics[width=1\linewidth]{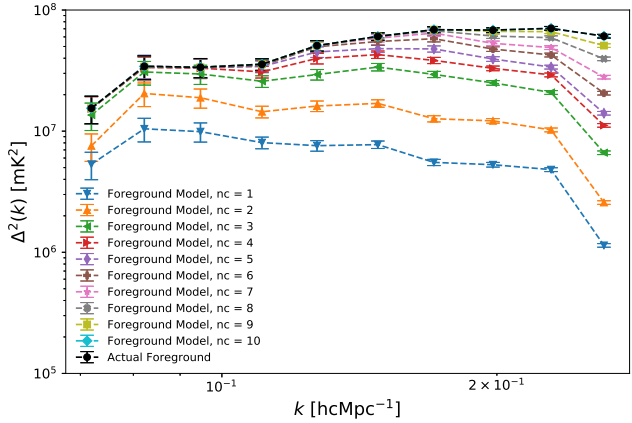}
\caption{A comparison of FastICA's foreground model power spectra, over bandwidth $\Delta \nu$ = 11.5 MHz. Using different numbers of components, ranging from 1 to 10.}
\label{fig:FG_REC}
\end{figure}

\begin{figure}
\centering
\includegraphics[width=1\linewidth]{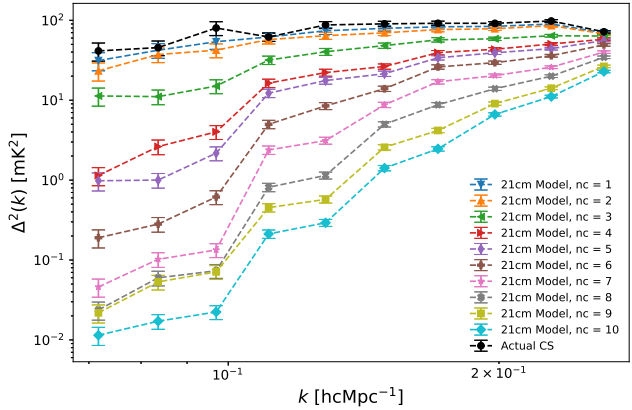}
\caption{Comparison of power spectra of FastICA's recovered simulated cosmological signal, over bandwidth $\Delta \nu$ = 11.5 MHz. Using different numbers of components, ranging from 1 to 10.}
\label{fig:CS_REC}
\end{figure}

BSS techniques are not scale-dependent, they apply the same mixing matrix model to all scales. On smaller \textit{k}-scales (larger scales in real space), there are a smaller number of independent components and so modelling the foregrounds with 6 components is sufficient. As we go to larger \textit{k}-scales (smaller scales in real space), the number of independent components increase, and so modelling the foregrounds with 6 components is no longer sufficient. As the BSS techniques \textit{do not} account for this scale-dependence, one sees an increasing under-performance of the techniques with \textit{k}. In Fig. \ref{fig:FG_REC}, we apply FastICA to \textit{only} our simulated foregrounds from Section \ref{Foregrounds}. 
From Fig. \ref{fig:FG_REC}, we see as we increase the number of components, with which we model the foregrounds, the better convergence the recovered power spectrum has with the power spectrum of the foreground we wish to model. We see that FastICA reaches convergence on lower scales, below  \textit{h}cMpc$^{-1}$, with just 3 components. This turn over of performance at  \textit{h}cMpc$^{-1}$, is what we also observe in Fig. \ref{fig:C007010_PS}. As for larger scales in Fig. \ref{fig:FG_REC}, as we increase the number of components, we see a progressively better fit at these scales. One can easily argue, from Fig. \ref{fig:FG_REC}, that using a higher number of components is better as we would get a better fit to the foregrounds. However, in Fig. \ref{fig:CS_REC}, we look at FastICA's recovery of the simulated cosmological signal. 
As we increase the number of components, we see that more of the simulated cosmological signal is modelled in the foreground model - $\textbf{As}$ in equation \eqref{eq:FASTICAMAIN}. With the smaller \textit{k}-scales being penalised more for the increase in components, than the larger \textit{k}-scales. We can also potentially see this over-subtraction/signal suppression at the lower \textit{k}-scales in Fig. \ref{fig:2D Ratio}'s first \textit{k$_\parallel$} bin. Such suppression by these techniques on these lower \textit{k}-scales is also seen by \citet{2020:CunningtonIrfanCarucci}.

From Figs. \ref{fig:FG_REC} and \ref{fig:CS_REC}, we can infer that, though a higher number of components will be able to model larger \textit{k}-scales, it will cause over-fitting of the signal at smaller \textit{k}-scales and results in the cosmological signal, on these smaller \textit{k}-scales, leaking into the foreground model. Therein lies the problematic nature of the scale-independence of these BSS techniques. If one were able to fit different scales with the components best suited for it, rather than having a signal number of components across all scales, we are more likely to get a better fit. In future work, we will explore techniques that are scale-dependent and compare their performance to GPR. The version of GMCA used here, is the default version used in \citet{UpperLim_HBA}. There have been significant improvements and advancements of this BSS\footnote{http://www.cosmostat.org/software/gmcalab}. For example, Local-Generalised Morphological Component Analysis (LGMCA)\footnote{https://www.cosmostat.org/software/lgmca} \citep{LGMCA}. LGMCA, allows the mixing matrix to vary across the pixels in an image.

\section{Discussion and Conclusion}
\label{discussion}

With the recent  2$\sigma$ upper limit of $\Delta^2_{21} <$ (73)$^2$ mK$^2$ at \textit{k} = 0.075 \textit{h}cMpc$^{-1}$ by \citet{MertensUpper2020} from 141 hours of data, and with LOFAR currently processing over 2000 hours of data from their observing run of the NCP, we aimed to compare removal techniques to assess how well they perform under different conditions and their limitations. In this paper, we attempt to replicate the analysis from \citet{MertensUpper2020}, applying FastICA, GMCA, and GPR to the data used in \citet{MertensUpper2020}, to see which performs better. We then replicate the data via simulation to pin down the reasoning as to why certain removal techniques perform worse than others. In Section \ref{result:LOFAR}, we can see we perfectly recover the upper limit found in \citet{MertensUpper2020} in our use of GPR. As we know more about the signals we wish to observe, plus what constitutes it (e.g. foregrounds and mode mixing), we can have more control over the characteristics and properties of the signal we wish to model/fit. This is where GPR becomes powerful, taking the properties of these constituents (e.g. coherency length) to better fit these foregrounds and recover the cosmological signal. Interestingly, we have found that FastICA and GMCA also reproduce this  2$\sigma$ upper limit. This is a powerful result as both GMCA and FastICA are blind to the data, i.e. knowing nothing a-priori about the data. 

From our results in Fig. \ref{fig:C007010_PS}, we see that on scales larger than  \textit{h}cMpc$^{-1}$ FastICA and GMCA, performing identically, begin to deviate away from the spectrum produced by GPR. We began looking at possible reasons for FastICA's and GMCA's under-performance at these scales by looking at instrumental effects. We simulate foregrounds and the cosmological signal at the same redshift and frequency range as the observed data. We use OSKAR to apply observational effects in three case, with noise included: 
\begin{enumerate}
  \item 120 hour observation.
  \item 120 hour observation, but we have removed the primary beam from the observation.
  \item 120 hour observation, but we have used the resolution from the lowest frequency in the observation’s point source function (PSF). For the purposes of this paper it is the PSF at 134.5 MHz; the primary beam is also included.
\end{enumerate}

From Figs. \ref{fig: Case FastICA} and \ref{fig:Case GMCA}, we find that the two  cases \textit{i} and \textit{ii} give similar power spectra. As we cut our FoV to be within the primary beam in the hope of mitigating most of the effects, this result is to be expected. As for case \textit{iii}, we see that though it gives a deeper power spectrum, there are clear signs of overfitting. We use the same number of components in all three cases to give a true comparison. It is clear here that the removal of chromaticity effects removes independent components from the data, especially at higher \textit{k}-scales. From this, we see it appears that no single instrumental effect is the dominant effect causing the under-performance at larger \textit{k}-scales - as we see no adverse effects on scales larger than  \textit{h}cMpc$^{-1}$ for both FastICA and GMCA. We then perform further checks that relate to the data itself. It is important to note how powerful the BSS techniques truly are. As we have said, they do not assume anything a-priori about the data. Yet, on simulated data, without OSKAR applied, they can recover the cosmological signal embedded in foregrounds.

We started our further analysis off with the amount of data we give our removal techniques. The original comparison performed in Fig. \ref{fig:Sim_Comp_Obs}, was done using a $\Delta \nu$ = 11.5 MHz data cube, between 134.5 MHz and 146 MHz, to match that of the \citet{MertensUpper2020} data. We then looked at two additional cube frequency bandwidths: $\Delta \nu$ = 30 MHz and $\Delta \nu$ = 50 MHz. We see in Fig. \ref{fig:Delta_nu_comp}, that there is an improvement in the results from FastICA and GMCA, with the $\Delta \nu$ = 30 MHz and $\Delta \nu$ = 50 MHz bandwidths performing better than the $\Delta \nu$ = 11.5 MHz case. However, there doesn't appear to be any significant improvement between $\Delta \nu$ = 30 MHz and $\Delta \nu$ = 50 MHz. Though we note for most applications of these BSS techniques, the bandwidth that is used is at least double of the highest we've given, $\Delta \nu$ = 50. For scales larger than  \textit{h}cMpc$^{-1}$, there is no answer to the under-performance - as we are seeing an improvement across all scales for the larger bandwidths. In the case of GPR, increasing the bandwidth worsens its performance. Larger bandwidths will average the frequency coherence of the data, and so one would expect GPR's performance to degrade with bandwidth. We have said that typically these BSS techniques require larger bandwidths, if GPR can perform well on relatively smaller bandwidths, it could prove advantageous. 

With no clear signs of degradation with the amount of data we use, we move on to see if noise was the limiting factor. We do not include noise in our OSKAR output and attempt to recover the cosmological signal. In our analysis, we limited the results to a fiducial value of 3 components, and did not see an improvement in the no noise case. This is interesting as the perfect noise avoidance of BSS techniques breaks down, as some noise must be in the foreground term. We see no improvement on scales larger than  \textit{h}cMpc$^{-1}$, and hence noise is not the limiting factor here. 

We next looked at the scale-independence of the BSS techniques. When applying FastICA or GMCA to data using a given number of components to model the data, it tries to model every scale of the data with that number of components. However, at smaller \textit{k}-scales there are a smaller number of independent components, and so modelling these scales with a small number of components is sufficient. As we move to larger \textit{k}-scales, the number of independent components also increases. Hence, the number of components that were sufficient to model the smaller \textit{k}-scales, is no longer sufficient at these larger \textit{k}-scales. We see this turning point at  \textit{h}cMpc$^{-1}$, in both Figs. \ref{fig:C007010_PS} and \ref{fig:FG_REC}. On scales below  \textit{h}cMpc$^{-1}$, using a handful of components to model the foregrounds is sufficient, but they are not sufficient to fully capture all the independent components at larger \textit{k}-scales. A simple approach would be to apply the larger number of independent components to model the foregrounds. However, all scales will be modelled with this number and we find that the 21cm signal loss is significant on smaller scales, when using this approach. When comparing Fig. \ref{fig:C007010_PS} to the results seen in Section \ref{result:Sim}, one can see that the number of components required to model the foregrounds is far greater, even on the smaller \textit{k}-scales, for the real data compared to the simulations. This is due to the short comings of simulations not being able to match excess power of higher \textit{k}-scales caused by, for example, RFI flagging or calibration errors. This then reduces the number of independent components seen at higher \textit{k}-scales, as such the scale-independence issue is not as significant and hence we see a deeper power spectrum at larger \textit{k}-scales for the simulated case versus the observed data. We also see the need of having a scale-dependent approach when looking to increase the bandwidth to improve the performance of BSS techniques. As aforementioned, applications of the BSS techniques have used larger bandwidths than that used in this paper as well as \citet{UpperLim_HBA} and \citet{LOFAR_UPPERLIMIT_GPR}. The applications to simulated data have been to that without instrumental effects, as such there is a linear improvement with bandwidth. Whereas for simulated case \textit{with} observational effects, we do see some improvements but it is likely that the scale-independence issue is more prominent and so the improvement isn't as clear. With real data, this is likely further exacerbated by the greater number of instrumental effects causing more independent components at high \textit{k}-scales.

If one were able to perform BSS on a scale-dependent approach, with a mixing matrix for different scales being modelled with different number of components, we believe we would likely see a better fit to the data and remove the divergence seen in Fig. \ref{fig:C007010_PS}. There is also no way to determine the best number of components to recover the signal, for each of these scales. One could, in future, aim to make these techniques less blind. As we have a grasp on the properties of the data, one could add a Bayesian back-end to these BSS techniques to optimise the number of components, on each scale, on which sparsity is performed.

Comparatively, GPR only takes into account the variance and optimises its fit with the knowledge of the constituents of the observed signal: that there is a weaker signal present, as it has a kernel modelling the signal, in the foregrounds and mode-mixing effects. It is less likely to overfit and remove power for the recovered signal, as it is accounted for, and is also able to model the mode mixing effect in the foregrounds as we optimise for its inclusion. GPR can also output the optimised covariance matrix, so one can see how it chose the fit - something neither FastICA or GMCA can do. We also see clear improvements in the performance of GPR as we increase the bandwidth over which we apply GPR, likely showing its accounting of mode-mixing makes instrumental effects less detrimental. 

GPR also has scope for improvement. GPR, currently, fits visibility by visibility, equivalent to one line-of-sight at a time, and so can be further improved by utilising all of the lines of sight together, to include spatial correlations, to better improve the fitting. As mentioned in Section \ref{GPR}, the covariance parameters for mode-mixing were based on treating mode-mixing as a Gaussian process rather than simulating the effect \citep{GPR,LOFAR_UPPERLIMIT_GPR}. If it were to be updated and tested on a more realistic mode-mixing, the covariance \textit{\textbf{K$_{\rm{fg}}$}} is more accurately tested and parameters refined, it will likely fit the data even better improving the recovery. GPR is also not scale-dependent, as the same covariance model for all visibilities. Improving this, is likely to yield a better recovery of the cosmological signal as well.
\newline

With LOFAR still processing the observed data from its observing run of the NCP, we conclude that the foreground removal technique that is best suited, out of the three tested methods, to recover the 21cm power spectrum by LOFAR is GPR, as used in \citet{MertensUpper2020}. Through the testing we have done in this paper that GPR has shown to be the most rigorous and still has scope to be improved on. Its application on the \citet{MertensUpper2020} data, looks to be close to noise-limited and as the full observational data of the NCP it processed, will likely detect the signal in this low-noise regime.

There is also need for further exploration of these BSS techniques in the production of upper limits. On \textit{k}-scales below  \textit{h}cMpc$^{-1}$, we have seen on these scales, on actual LOFAR data, they perform as well as GPR. As aforementioned, BSS techniques are very powerful as they do not assume anything about the data a-priori, and there have been many further iterations of GMCA. Application of a scale-dependent approach is something to be looked at in future works. Machine learning is likely going to play a prominent role in foreground mitigation in the near future, with a framework being set up for LOFAR-EoR. Sooknunan et al., in prep, are developing a machine learning framework for LOFAR-EoR foreground mitigation techniques.

We have also seen that \citet{MertensUpper2020} compared to \citet{UpperLim_HBA}, as well as using a different foreground removal technique, has improved the direction-dependent calibration of the data. Once LOFAR has fully processed the data from its observing run of the NCP, it is likely to have significantly improved its calibration, as well as improving the current sky model and the inclusion of phase errors introduced by the ionosphere. This will yield deeper power spectra limits as well as increasing the likelihood of detecting the cosmological signal with the fully processed data.

\section*{Acknowledgements}
EC acknowledges the support of a Royal Society Dorothy Hodgkin Fellowship and a Royal Society Enhancement Award. JRP acknowledges financial support from the European Research Council under ERC grant number 638743-FIRSTDAWN. FGM and LVEK would like to acknowledge support from a SKA-NL Roadmap grant from the Dutch ministry of OCW. ITI was supported by the Science and Technology Facilities Council [grant numbers ST/I000976/1 and ST/T000473/1] and the Southeast Physics Network (SEP-Net). VJ acknowledges support by the Croatian Science Foundation for the project IP-2018-01-2889 (LowFreqCRO). B.K.G. would like to acknowledge the National Science Foundation grant AST-1836019. We also thank Catherine A. Watkinson for her useful discussions and comments. This research made use of astropy, a community-developed core Python package for astronomy \citep{Astropy}; scipy, a Python-based ecosystem of open-source software for mathematics, science, and engineering \citep{Scipy} - including numpy \citep{numpy}; matplotlib, a Python library for publication quality graphics \citep{matplotlib}; scikit-learn, a free software machine learning library for the Python programming language, for the use of FastICA \citep{scikit-learn}.

\section*{Data availability}

The data underlying this article were provided by LOFAR under licence / by permission. Data will be shared on request to the corresponding author with permission of LOFAR.



\bibliography{cit}

\bsp	
\label{lastpage}
\end{document}